\documentclass[aps,pre,twocolumn,superscriptaddress]{revtex4-1}
\usepackage{color,dsfont}
\usepackage{amsmath, soul, xcolor}
\usepackage{url}
\usepackage{array}
\usepackage{graphicx} 
\usepackage{bmpsize}
\usepackage{amssymb}

\begin{document}


\title{Multiplexity and multireciprocity in directed multiplexes}


\author{Valerio Gemmetto}
\affiliation{Instituut-Lorentz for Theoretical Physics, Leiden Institute of Physics, University of Leiden, The Netherlands}

\author{Tiziano Squartini}
\affiliation{IMT Institute for Advanced Studies, Lucca, Italy}

\author{Francesco Picciolo}
\affiliation{Department of Biotechnology, Chemistry and Pharmacy, University of Siena, Italy}

\author{Franco Ruzzenenti}
\affiliation{Department of Economics and Statistics, University of Siena, Italy}
\affiliation{Institute of Sociology, Jagiellonian University, Krakow, Poland}
\affiliation{Department of Management, University of Naples "Parthenope", Naples, Italy}

\author{Diego Garlaschelli}
\affiliation{Instituut-Lorentz for Theoretical Physics, Leiden Institute of Physics, University of Leiden, The Netherlands}


\date{\today}

\begin{abstract}
Real-world multi-layer networks feature nontrivial dependencies among links of different layers. Here we argue that, if links are directed, dependencies are twofold. Besides the ordinary tendency of links of different layers to align as the result of `multiplexity', there is also a tendency to anti-align as the result of what we call `multireciprocity', i.e. the fact that links in one layer can be reciprocated by \emph{opposite} links in a different layer. 
Multireciprocity generalizes the scalar definition of single-layer reciprocity to that of a square matrix involving all pairs of layers. We introduce multiplexity and multireciprocity matrices for both binary and weighted multiplexes and validate their statistical significance against maximum-entropy null models that filter out the effects of node heterogeneity.
We then perform a detailed empirical analysis of the World Trade Multiplex (WTM), representing the import-export relationships between world countries in different commodities. We show that the WTM exhibits strong multiplexity and multireciprocity, an effect which is however largely encoded into the degree or strength sequences of individual layers. 
The residual effects are still significant and allow to classify pairs of commodities according to their tendency to be traded together in the same direction and/or in opposite ones.
We also find that the multireciprocity of the WTM is significantly lower than the usual reciprocity measured on the aggregate network. Moreover, layers with low (high) internal reciprocity are embedded within sets of layers with comparably low (high) mutual multireciprocity. This suggests that, in the WTM, reciprocity is inherent to groups of related commodities rather than to individual commodities. 
We  discuss the implications for international trade research focusing on product taxonomies, the product space, and fitness/complexity metrics.
\end{abstract}


\maketitle

\section{Introduction}
Several real-world systems are composed by intricately interconnected units, thus exhibiting a nontrivial network structure. 
The behaviour and dynamics of such systems are strongly dependent on how information can propagate throughout the network. 
Both the directionality and the intensity of connections crucially affect this process, and should possibly be incorporated in the network description.
For instance, most of the communication relations among individuals, such as exchanges of letters, e-mails or texts, are intrinsically directional and are therefore best represented as directed networks~\cite{Ebel}.
Furthermore, such interactions typically have heterogeneous intensities, calling for a description in terms of weighted networks~\cite{Newman}.\\

Recently, it has been realized many real-world systems often require an even more detailed representation, because a given set of units can be connected by different kinds of relations.
This property can be abstractly captured in terms of so-called \emph{edge-colored graphs} (where links of different colors are allowed among the same set of nodes) or equivalenlty \emph{multi-layer} or \emph{multiplex networks} (where the same set of nodes is replicated in multiple layers, each of which is an ordinary network)~\cite{Boccaletti, Kivela}.
The nontrivial properties of these systems, with respect to ordinary single-layer (`monochromatic' or `monoplex') networks, arise from the facts that the various layers are interdependent and the presence of a link in one layer can influence the presence of a link in a different layer. 
A clear example is represented by the different 
kinds of relationships existing between employees in a university department~\cite{Magnani}, where individuals can be connected by co-authorship, common leisure activities, on-line social networks etc.
The interdependence of layers implies that the topological properties usually defined for monoplex networks admit nontrivial generalizations to multiplex networks, and that some properties which are uninteresting, or even undefined, for single-layer networks become relevant for multiplexes.\\

This paper introduces novel metrics characterizing the dependencies among layers in multiplexes \emph{with directed links}. 
While various measures of inter-layer overlap for multiplexes have already been introduced \cite{Battiston, Bianconi}, they suffer from two main limitations.
First, most definitions are available only for multiplexes with undirected links, and their straightforward generalization to the directed case would overshadow important properties that are inherent to directed networks, most importantly the reciprocity (which is one of our main focuses here). 
Second, even in `trivial' multiplexes where there is no dependence among layers (i.e. in independent superpositions of single-layer networks with the same set of nodes), a certain degree of inter-layer overlap can be created entirely by chance. 
This effect becomes more pronounced as the density of the single-layer networks increases and as the correlation among single-node properties (like degrees or strengths) across layers increases. 
For instance, if a node is a hub in multiple layers, there is an increased chance of overlap among these layers, even if the presence of links in one layer is assumed not to influence the presence of links in another layer.\\

The two limitations discussed above highlight the need to define metrics that appropriately filter out both global (network-wide) and local (node-specific) density effects. 
Correlation-based measures of inter-layer overlap have been proposed with this aim in mind \cite{barigozzi}.
However, as recently pointed out \cite{gemmetto}, correlation-based metrics for multiplexes are not a correct solution in general, because they implicitly assume that edges observed between different pairs of nodes are sampled from the same probability distribution. This assumption is strongly violated in real-world networks, whose markedly heterogeneous topology is a signature of very different probabilities for edges emanating from different nodes, e.g. the probability of links being found around more important nodes is clearly different from the probability of links being found around less important nodes.\\

The above considerations motivate us to introduce new multiplexity metrics that explicitly take the directionality of links into account and appropriately filter out the spurious effects of chance, while controlling for the extreme heterogeneity of empirical node-specific properties.  
In this paper we carry out this program by extending recent `filtered' definitions of multiplexity \cite{gemmetto}, originally defined for undirected links, to the case of directed links. Although this might seem a straightforward procedure at first, we will in fact show that it requires different null models, triggers novel concepts, and leads to new quantities that are undefined in the undirected case. 
Indeed, while in the undirected case there is only one possible notion of dependency among links in different layers, in the directed case there are two possibilities, depending on whether links are `aligned' or `anti-aligned'.\\

Aligned links between two layers are observed when a directed link from node $i$ to node $j$ exists in both layers. This situation is the straightforward analogue of what can happen in undirected multiplex networks, and is a signature of the fact that the connection from $i$ to $j$ is relevant for multiple layers. 
We will denote this effect simply as \emph{multiplexity}, in analogy with the undirected case \cite{gemmetto}, and will study it in the general case of an arbitrary number of layers.
By contrast, anti-aligned links form between two layers when a link from node $i$ to node $j$ in one layer is \emph{reciprocated} by an opposite link from node $j$ to node $i$ in the other layer.
This situation does not have a counterpart in the case of undirected multiplexes and leads us to the definition of the novel concept of \emph{multireciprocity}, i.e. the generalization of the popular concept of reciprocity to the case of multiplex networks.\\

In monoplex networks - either binary~\cite{Garlaschelli2} or weighted~\cite{Squartini} - reciprocity is defined as the tendency of vertex pairs to form mutual connections. 
This property, which is one of the best studied properties of single-layer directed networks, can crucially affect various dynamical processes such as diffusion~\cite{Meyers}, percolation~\cite{Boguna} 
and growth~\cite{Schnegg, Garlaschelli1}. 
For instance, the presence of directed, reciprocal connections can lead to 
the establishment of functional communities and hierarchies of groups of neurons in the cerebral cortex~\cite{Zhou}.\\

In binary graphs, a simple measure of reciprocity is the ratio of the number of reciprocated links (i.e. realized links for which the link pointing in the opposite direction between the same two nodes is also realized) to the total number of directed links. 
However, it has been shown~\cite{Garlaschelli2} that this measure is not \emph{per se} informative about the actual tendency towards reciprocation, because even in a random network a certain number of reciprocated links will appear. 
So the number of observed mutual interactions has to be compared with the expected number obtained for a given random null model, if one wants to understand whether mutual links are present in the real network significantly more (or less) often than 
in the random benchmark~\cite{Newman2}.
It is therefore crucial to make use of proper null models for networks. 
Since in most real-world directed networks the distribution of the number of in-coming and out-going links (i.e. the in-degree and out-degree) of nodes is very broad, an appropriate null model should fix the in- and out-degrees of all nodes equal to their observed values. The null model of directed networks with given in- and out-degrees often goes under the name of directed binary configuration model (DBCM)~\cite{Maslov}. The rationale underlying the DBCM is the consideration that the in- and out-degree of a node might reflect some intrinsic `size', or other characteristic, of that node; therefore a null model tailored for a specific network should preserve the observed degree heterogeneity. 
Conveniently, the DBCM is also the correct null model to use when measuring the \emph{multiplexity} among layers of a multiplex with directed links.
Indeed, the DBCM is the directed generalization of the undirected binary configuration model used in \cite{gemmetto} for the definition of appropriately filtered, undirected multiplexity metrics. 
This nicely implies that we can use the DBCM as a single null model in our analysis of both multiplexity and multireciprocity.\\

Recently, the definition of reciprocity has been extended to weighted networks~\cite{Squartini}. A simple measure of weighted reciprocity is the ratio of `total reciprocated link weight' to total link weight, where the reciprocated link weight is defined, for any two reciprocated links, as the minimum weight of the two links. 
Similarly to the binary case, some level of weighted reciprocity can be generated purely by chance. So the empirical measure has to be compared to its expected value under a proper null model, represented in this case by a random weighted network where each node has the same in-strength and out-strength (i.e. total in-coming link weight and total out-going link weight, respectively) as in the real network. This null model is sometimes called the directed weighted configuration model (DWCM)~\cite{Serrano} and, conveniently, is also the relevant null model (generalizing its undirected counterpart \cite{gemmetto}) to study the multiplexity in presence of weighted directed links.\\

We stress that the concept of reciprocity has not been generalized to multiplex networks yet. Our definition of multireciprocity represents the first step in this direction and captures the tendency of a directed link in one layer of a multiplex to be reciprocated by an opposite link in a possibly \emph{different} layer. While ordinary reciprocity can be quantified by a scalar quantity, multireciprocity requires a square matrix where all the possible pairs of layers are considered.
Similarly, the multiplexity also requires a square matrix. 
Together, the multiplexity matrix and the multireciprocity matrix represent the two `directed' extensions of the undirected multiplexity matrix that has recently been introduced~\cite{gemmetto} to characterize undirected (either binary or weighted) multiplexes.\\

The rest of the paper is organized as follows. In Sec.~\ref{sec:defs} we introduce our methods, null models and main definitions for both binary and weighted multiplexes. In Sec.~\ref{sec:WTM} we apply our techniques to the analysis of the World Trade Multiplex (WTM), a directed weighted multiplex representing the import-export relations between countries of the world in different products. We identify a number of empirical properties of the WTM that are impossible to access via the usual aggregate (monoplex) analysis of the network of total international trade.
We finally conclude the paper in Sec.~\ref{sec:conclusions}, where we discuss some important implications of our results, both for the general study of multiplex networks and for more specific research questions in international trade economics. Several necessary technical details are given in the Appendices.

\section{Multiplexity and Multireciprocity metrics\label{sec:defs}}
In this section we give definitions of multiplexity and multireciprocity metrics for both binary and weighted multiplexes. These definitions require, as a preliminary step, the introduction of appropriate null models.
In turn, null models require the choice of a convenient notation.
We address these points in the resulting order.

We represent a directed multiplex $ \overrightarrow{G} = (G^1,\ldots,G^M) $ as the superposition of $ M $ directed networks (layers) $ G^{\alpha} $ ($ \alpha = 1,\ldots,M $), all sharing the same set of $N$ nodes~\cite{Boccaletti}. 
Links can be either binary or weighted.
In the binary case, each layer $\alpha$ is represented by a $N\times N$ binary adjacency matrix $G^{\alpha}=(a_{ij}^\alpha)_{i,j=1}^N$, where $a_{ij}^\alpha=0,1$ depending on whether a directed link from node $i$ to node $j$ is absent or present, respectively.
In the weighted case, each layer $\alpha$ is represented by a $N\times N$ non-negative integer adjacency matrix $G^{\alpha}=(w_{ij}^\alpha)_{i,j=1}^N$, where $w_{ij}^\alpha=0,1,\dots\infty$ is the weight of the directed link from node $i$ to node $j$ ($w_{ij}^\alpha=0$ indicating the absence of such link).
We denote by $\mathcal{G}_N$ the set of all (binary or weighted) single-layer graphs with $N$ nodes, and by $\mathcal{G}_N^M\equiv(\mathcal{G}_N)^M $ the set of all (binary or weighted) $M$-layer multiplexes with $N$ nodes.

\subsection{Null models of multiplex networks: maximum entropy and maximum likelihood}

Since our purpose is that of measuring correlations between directed links (possibly, in opposite directions) in different layers, we define independent reference 
models for each layer of the multiplex, thus creating an uncorrelated null model for the entire multiplex~\cite{gemmetto,Bianconi}.
This means that, if $ \mathcal{P}(\overrightarrow{G}|\overrightarrow{\theta})$ denotes the joint probability of the entire multiplex $\overrightarrow{G}\in\mathcal{G}_N^M$ (given a set of constraints enforced via the vector $\overrightarrow{\theta}$ of parameters, see Appendix \ref{app:maxent}) and 
\begin{equation}
P^{\alpha}(G^{\alpha}|\overrightarrow{\theta^{\alpha}})\equiv
\underbrace{\sum_{G^1\in \mathcal{G}_N}\dots\sum_{G^\beta\in \mathcal{G}_N}\dots\sum_{G^M\in \mathcal{G}_N}}_{\beta\ne\alpha}
\mathcal{P}(\overrightarrow{G}|\overrightarrow{\theta})
\label{eq:marginal}
\end{equation}
denotes the (marginal) probability for the single-layer graph $G^{\alpha}\in\mathcal{G}_N$ (given a set of layer-specific constraints enforced via the partial vector $\overrightarrow{\theta^{\alpha}}$), we require the null model to obey the factorization property
\begin{eqnarray}
\mathcal{P} \big(\overrightarrow{G} |\overrightarrow{\theta}\big) = \prod_{\alpha = 1}^{M} P^{\alpha}(G^{\alpha}|\overrightarrow{\theta^{\alpha}}).
\label{prob_multiplex}
\end{eqnarray}
The above property ensures that the definition of the null model for the entire multiplex reduces to the definition of independent null models for each layer separately (see Appendix~\ref{app:maxent} for a rigorous derivation).\\

In the case of binary multiplexes, the null model we want to use to control for the heterogeneity of nodes in each layer is, as we have already mentioned, the Directed Binary Configuration Model (DBCM)~\cite{Squartini1,Squartini6}, defined as the ensemble of binary networks with given in-degree and out-degree sequences.
At this point, we have to make a major decision, since the DBCM can be implemented either microcanonically or canonically.\\

In the microcanonical approach, node degrees are ``hard'', i.e. enforced sharply on each realization. The most popular microcanonical implementation of the DBCM is based on the random degree-preserving rewiring of links~\cite{Maslov} (a.k.a. the Local Rewiring Algorithm), which unfortunately introduces a bias. This bias arises because, if the degree distribution is sufficiently broad (as in most real-world cases), the randomization process explores the space of possible network configurations not uniformly, giving higher probability to the configurations that are ``closer'' to the initial one~\cite{Roberts} (more details are given in Appendix~\ref{app:maxlike}).
Another possible microcanonical implementation, based on the random matching of ``edge stubs'' (half links) to the nodes, creates undesired self-loops and multiple edges~\cite{Maslov,Newman1}.
Besides these limitations, microcanonical approaches are computationally demanding. Indeed, in order to measure the expected value of any quantity of interest, it is necessary to generate several randomized networks, on each of which the quantity needs to be calculated. This sampling method is \textit{per se} very costly, and even more so in the case of multiplex networks, due to the presence of several layers requiring a further multiplication of iterations (see Appendix~\ref{app:maxlike}).\\

By contrast, in the canonical implementation~\cite{Squartini1,Squartini6} of the DBCM the in- and out-degrees are ``soft'', i.e. preserved only on average. The resulting probability distribution over the ensemble of possible graphs is obtained analytically by maximizing the entropy subject to the enforced constraints~\cite{Robins,Park,Park1,Squartini1} (see Appendix~\ref{app:maxent} for details). This procedure leads to the class of models also known as Exponential Random Graphs or $p^{\star}$ models~\cite{Holland,Wasserman,Snijders}. 
In order to fit such exponential random graphs to real-world networks, we adopt an exact, unbiased and fast method~\cite{Squartini1,Squartini6} based on the Maximum Likelihood principle~\cite{Garlaschelli0}. The method is summarized in Appendix~\ref{app:maxlike} and implemented in our analysis using the so-called MAX$\&$SAM (``Maximize and Sample'') algorithm~\cite{Squartini6}. 
The latter yields the exact probabilities of occurrence of any graph in the ensemble and the explicit expectation values of 
the quantities of interest. 
This has the enormous advantage that an explicit sampling of graphs is not required: expectation values are calculated analytically and not as sample averages.
In particular, the probability $ p_{ij}^{\alpha} $ that a link from node $ i $ to node $ j $ is realized in layer $ \alpha $ ($a_{ij}^\alpha=1$) can be easily calculated. From the set of all such probabilities, the expected value of the multireciprocity can be computed analytically and directly compared with the empirical value, in order to obtain a filtered measure.\\

We now come to the case of multiplexes with weighted links.
In this case we want the enforced constraints to be the in-strength and out-strength sequences of the real network, separately for each layer. 
The corresponding model is sometimes referred to as the Directed Weighted Configuration Model (DWCM)~\cite{Squartini}. 
As for the binary case, we want to build the null model canonically as a maximum-entropy ensemble of weighted networks, leading to a weighted Exponential Random Graph model~\cite{Squartini1,Squartini}. 
The implementation we use is again based on the MAX$\&$SAM algorithm~\cite{Squartini6}, which in this case calculates the exact probability that, in the null model, the weight of the directed link connecting node $ i $ to node $ j $ in layer $ \alpha $ has a particular value $w_{ij}^\alpha$, for each pair of nodes and each layer.
From this probability, the expected weighted multireciprocity can be computed analytically and compared with the empirical one, thus producing a filtered value that, in this case as well, does not require the explicit sampling of graphs.

\subsection{Binary multiplexity and multireciprocity}
Our first set of main definitions are specific for multiplexes with binary links.
Consider a directed and binary multiplex $\vec{G}$ with $M$ layers. 
We quantify the similarity and reciprocity between any two layers $\alpha$ and $\beta$ by defining the binary \emph{multiplexity} $m^{\alpha, \beta}_b$ and \emph{multireciprocity} $r^{\alpha, \beta}_b$ as follows:
\begin{subequations}
\begin{align}
m^{\alpha, \beta}_b= \frac{2 \sum_i\sum_{j \neq i} \min \{ a_{ij}^{\alpha}, a_{ij}^{\beta}\}}{L^{\alpha} + L^{\beta}}= \frac{2 L^{\alpha\rightrightarrows\beta}}{L^{\alpha} + L^{\beta}}, &\\ 
r^{\alpha, \beta}_b = \frac{2 \sum_i\sum_{j \neq i} \min \{ a_{ij}^{\alpha}, a_{ji}^{\beta}\}}{L^{\alpha} + L^{\beta}}= \frac{2 L^{\alpha\rightleftarrows\beta}}{L^{\alpha} + L^{\beta}}, &
\end{align}
\label{m_bin}
\end{subequations}
\noindent where $ L^{\alpha}=\sum_i\sum_{j \neq i}a^\alpha_{ij} $ 
represents the total number of directed links in layer $ \alpha $ (analogously for layer $ \beta $), $ L^{\alpha\rightrightarrows\beta}=\sum_i\sum_{j \neq i} \min \{ a_{ij}^{\alpha}, a_{ij}^{\beta}\}$ is the number of links of layer $\alpha$ that are multiplexed in layer $\beta$ (clearly, $ L^{\alpha\rightrightarrows\beta}= L^{\beta\rightrightarrows\alpha}$), and $ L^{\alpha\rightleftarrows\beta}=\sum_i\sum_{j \neq i} \min \{ a_{ij}^{\alpha}, a_{ji}^{\beta}\}$ is the number of links of layer $\alpha$ that are reciprocated in layer $\beta$ (clearly, $ L^{\alpha\rightleftarrows\beta}= L^{\beta\rightleftarrows\alpha}$).  
Note that possible self-loops (terms of the type $a_{ii}^\alpha$) are deliberately ignored because they are indistinguishable from links pointing in the opposite direction, thus making their contribution to either multiplexity or multireciprocity undefined.

Equations \eqref{m_bin} can be regarded as defining the entries of two $M\times M$ matrices, which we will call the \emph{binary multiplexity matrix} ${\bf M}_b=(m^{\alpha, \beta}_b)_{\alpha,\beta}$ and the \emph{binary multireciprocity matrix} ${\bf R}_b=(r^{\alpha, \beta}_b)_{\alpha,\beta}$ respectively.
The matrices ${\bf M}_b$ and ${\bf R}_b$ represent the two natural extensions, to the case of directed multiplexes, of the single binary multiplexity matrix introduced in~\cite{gemmetto} for undirected binary multiplexes.
Both matrices provide information about the `overlap' between directed links connecting pairs of nodes in different layers. 
Their entries range in $ [0,1] $ and are maximal only when layers $ \alpha $ and $ \beta $ are respectively identical (i.e. $a_{ij}^\alpha=a_{ij}^\beta$ for all $i\ne j$) and fully `multireciprocated' (i.e. $a_{ij}^\alpha=a_{ji}^\beta$ for all $i\ne j$).
The matrix ${\bf M}_b$ has by construction a unit diagonal, since the intra-layer multiplexity trivially has the maximum value $ m^{\alpha, \alpha}_b = 1 $ for all $\alpha$. By contrast, the diagonal of ${\bf R}_b$ is nontrivial and of special significance, as the intra-layer multireciprocity $ r^{\alpha, \alpha}_b $ reduces to the ordinary definition of binary reciprocity for monoplex networks~\cite{Garlaschelli2}.\\

For `trivial', uncorrelated multiplexes made of sparse non-interacting layers with narrow degree distributions, the matrix ${\bf M}_b$ would asymptotically (i.e. in the limit of large $N$, but not necessarily large $M$) be the $M\times M$ identity matrix, and the matrix ${\bf R}_b$ would asymptotically be a $M\times M$ diagonal matrix. This is because, in presence of sparse uncorrelated layers without hubs, the chance of a link in one layer `overlapping' with a (mutual) link in a different layer is negligible. 
For finite and/or dense networks and/or broad degree distributions, however, positive values of $m^{\alpha, \beta}_b$ and $r^{\alpha, \beta}_b$ (with $\alpha\ne\beta$) can be produced entirely by chance even in a multiplex with no dependencies among layers.
For instance, if the same node is a hub in multiple layers, the chance of a large overlap of links among all pairs of such layers is very high, even if the layers are non-interacting.\\

The above considerations imply that, in order to extract statistically significant information about the tendency towards multiplexity and multireciprocity in a real-world multiplex, it becomes necessary to compare the empirical values of $m^{\alpha, \beta}_b$ and $r^{\alpha, \beta}_b$ with the corresponding expected values calculated under the chosen null model of independent multiplexes with given degrees (i.e. the DBCM). 
Hence, we introduce the \emph{transformed} binary multiplexity and  
multireciprocity matrices with entries
\begin{subequations}
\begin{align}
\mu^{\alpha, \beta}_b = \frac{m^{\alpha, \beta}_b - \langle m^{\alpha, \beta}_b \rangle_{\rm DBCM}}{1 - \langle m^{\alpha, \beta}_b \rangle_{\rm DBCM}}&\qquad(\alpha\ne\beta), \\ 
\rho^{\alpha, \beta}_b = \frac{r^{\alpha, \beta}_b- \langle r^{\alpha, \beta}_b \rangle_{\rm DBCM}}{1 - \langle r^{\alpha, \beta}_b \rangle_{\rm DBCM}},&
\end{align}
\label{def_mu_bin}
\end{subequations}
where $ \langle\cdot \rangle_{\rm DBCM}$ denotes the expected value under the DBCM. Note that, since $\langle m^{\alpha, \alpha}_b\rangle_{DBCM}=m^{\alpha, \alpha}_b=1$ for all $\alpha$, we formally set the diagonal terms $\mu^{\alpha, \alpha}_b\equiv 1$, as the definition (\ref{def_mu_bin}a) would produce an indeterminate expression if extended to $\alpha=\beta$.
The explicit calculation of the above expected values is provided in Appendix~\ref{app:DBCM} and more details are provided later in this section.\\

The filtered quantities \eqref{def_mu_bin} are directly informative about the presence of dependencies between layers. Positive values represent higher-than-expected multiplexity or multireciprocity (correlated or `attractive' pairs of layers), while negative values represent lower-than-expected quantities (anticorrelated or `repulsive' pairs of layers). Pairs of uncorrelated (`noninteracting') layers are characterized by multiplexity and multireciprocity values comparable with 0. In principle, a layer that is uncorrelated with all other layers can be separated from the multiplex and analysed separately from it.\\

The choice of the denominator of~(\ref{def_mu_bin}a) and~(\ref{def_mu_bin}b), \emph{a priori} not obvious, guarantees that the maximum value for the transformed multiplexity and multireciprocity is 1. Moreover, it ensures that $ \rho^{\alpha, \alpha}_b $ reduces to the rescaled reciprocity $\rho_b $ defined for single-layer networks~\cite{Garlaschelli2}.
It should also be noted that the multiplexity defined in (\ref{m_bin}a) is just the normalized version of the inter-layer overlap introduced in~\cite{Battiston} and~\cite{Bianconi}, extended to directed multiplex networks. In this context, the novel contribution that we give is the comparison with a null model. Indeed, while (\ref{m_bin}a)  only provides information about the raw similarity of the layers, which is strongly density-dependent, the transformed measure (\ref{def_mu_bin}a)  is mapped to a universal interval. In combination with the $z$-scores that we introduce later, it can be used to consistently compare the statistical significance of the multiplexity of different systems. The quantity defined in (\ref{m_bin}b), which focuses explicitly on the reciprocity properties of the multiplex, has never been introduced before, along with its transformed quantity defined in (\ref{def_mu_bin}b). The latter can be used for a consistent comparison of the multireciprocity of mutiplexes with different densities.\\

The calculation of the expected values of $m^{\alpha, \beta}_b$ and $r^{\alpha, \beta}_b$ under the DBCM can be carried out analytically using the MAX$\&$SAM method~\cite{Squartini6}, with no need to actually randomize the empirical network or numerically sample the null model ensemble.  
Ultimately, the calculation requires the computation of the expected value of the minimum between two binary random variables (see Appendix~\ref{app:DBCM}). If $p_{ij}^{\alpha}\equiv \langle a_{ij}^{\alpha} \rangle_{\rm DBCM}$ denotes the probability that, under the DBCM, a directed link is realized from node $ i $ to node $ j $ in layer $\alpha$, then the adjacency matrix entry $a_{ij}^{\alpha}$ is described by the Bernoulli mass probability function 
\begin{eqnarray}
P(a_{ij}^{\alpha}) = (p_{ij}^{\alpha})^{a_{ij}^{\alpha}} (1 - p_{ij}^{\alpha})^{(1 - a_{ij}^{\alpha})}.
\label{eq:Bernoulli}
\end{eqnarray}
Using the above equation, and given the explicit expression for $p_{ij}^\alpha$, it is possible to calculate $ \mu^{\alpha, \beta}_b $ and $ \rho^{\alpha, \beta}_b $ analytically as reported in Appendix~\ref{app:DBCM}.\\

It is instructive to compare the multivariate quantities measured on the multiplex with the corresponding scalar quantities defined on the aggregate monoplex network obtained by combining all layers together. This comparison can highlight the gain of information resulting from the multiplex representation, with respect to the ordinary monoplex projection where all the distinct types of links are treated as equivalent. 
The binary aggregate monoplex can be defined in terms of the adjacency matrix with entries
\begin{equation}
a_{ij}^{\rm mono} =1-\prod_{\alpha=1}^M(1-a_{ij}^\alpha)= \left\{
    \begin{array}{ll}
      1 & \textrm{if  } \exists \alpha : a_{ij}^{\alpha} = 1 \\
      0 & \mbox{otherwise}
     \end{array}
  \right. .
\label{mono_b}
\end{equation}
For the quantities we defined so far, the only meaningful comparison between the multiplex and the aggregate network can be done in terms of the reciprocity, because the multiplexity of the aggregate is $m^{\rm mono}_b =1$ by construction.
The single, global reciprocity of the aggregated monoplex network is given by
\begin{equation}
r^{\rm mono}_b = \frac{\sum_i\sum_{j \neq i} \min \{ a_{ij}^{\rm mono}, a_{ji}^{\rm mono}\}}{L^{\rm mono}}
\label{r_bin_aggr}
\end{equation}
\noindent where $ L^{\rm mono}=\sum_i\sum_{j \neq i}a^{\rm mono}_{ij} $. Similarly, it is possible to define the corresponding filtered quantity $\rho^{\rm mono}_b$, in analogy with~(\ref{def_mu_bin}b).\\

The transformed quantities $\mu^{\alpha, \beta}_b$ and $\rho^{\alpha, \beta}_b$ defined in~(\ref{def_mu_bin}) capture the similarity and reciprocity between layers of a multiplex via a comparison of the empirical values with the expected values under a null model. However, those quantities do not consider any information about the variances of the values of multiplexity and multireciprocity under the null model, thus giving no direct information about statistical significance. 
In particular, even multiplexes sampled from the null model with independent layers would be characterized by small, but in general nonzero, values of  $\mu^{\alpha, \beta}_b$ and $\rho^{\alpha, \beta}_b$.
This makes it difficult to disentangle, for an observed real-world multiplex, weak inter-layer dependencies from pure noise.
Moreover, the random fluctuations around the expectation values will be in general different for different pairs of layers, potentially making the comparison of the values of $\mu^{\alpha, \beta}_b$ and $\rho^{\alpha, \beta}_b$ for different pairs of layers misleading.
To overcome these limitations, we define the $z$-scores associated to $ m^{\alpha, \beta}_b $ and $ r^{\alpha, \beta}_b $ as:
\begin{subequations}
\begin{align}
z \big( m^{\alpha, \beta}_b \big) = \frac{m^{\alpha, \beta}_b - \langle m^{\alpha, \beta}_b \rangle_{\rm DBCM}}{\sqrt{\langle ( m^{\alpha, \beta}_b )^2 \rangle_{\rm DBCM} - \langle m^{\alpha, \beta}_b \rangle_{\rm DBCM}^2}}, & \\ 
z \big( r^{\alpha, \beta}_b \big) = \frac{r^{\alpha, \beta}_b - \langle r^{\alpha, \beta}_b \rangle_{\rm DBCM}}{\sqrt{\langle ( r^{\alpha, \beta}_b )^2 \rangle_{\rm DBCM} - \langle r^{\alpha, \beta}_b \rangle_{\rm DBCM}^2}}. &
\end{align}
\label{zScores}
\end{subequations}
As for the quantities defined in (\ref{def_mu_bin}), it is possible to obtain an analytical expression for the $z$-scores as well. This is shown in detail in Appendix~\ref{app:DBCM}.\\

Each $z$-score in (\ref{zScores}) has the same sign as the corresponding quantity in (\ref{def_mu_bin}), since the numerator is the same and both have positive denominators. However, except for the common sign, the two sets of quantities can have \emph{a priori} very different values. 
In particular, the $z$-scores count the number of standard deviations by which the observed raw quantities deviate from their expected values under the null model. As such, they are useful in order to understand whether small measured values of $ \mu^{\alpha, \beta}_b $ or $ \rho^{\alpha, \beta}_b $ are actually consistent with zero within a small number of standard deviations, in which case we can consider the layers $ \alpha $ and $ \beta $ as uncorrelated.
We point out that, in general, $z$-scores have a clear statistical interpretation only if their distribution is Gaussian under repeated realizations of the model. In our case, although the quantities $ m^{\alpha, \beta}_b $ and $ r^{\alpha, \beta}_b $ are not truly normally distributed under the null model, they are defined as the sum of many independent 0/1 random variables (of the type $\min \{ a_{ij}^{\alpha}, a_{ij}^{\beta}\}$ or $\min \{ a_{ij}^{\alpha}, a_{ji}^{\beta}\}$ respectively), which all have variance in the interval $(0,1/4]$ and are thus approximately described by a central limit theorem ensuring an asymptotic convergence to the normal distribution. 
We can therefore consider as statistically significant all the $z$-scores having an absolute value larger than a given threshold, which we set at $z_{c}= 2 $. This selects the observed pairs of layers with values of multiplexity and/or multireciprocity that differ from their expectation values by more than 2 standard deviations, i.e. with $|z|>z_c$.

\subsection{Weighted multiplexity and multireciprocity}

We now move to our second set of definitions, valid for weighted multiplexes.
In analogy with (\ref{m_bin}), we define the \emph{weighted} multiplexity and multireciprocity matrices ${\bf M}_w$ and ${\bf R}_w$ having entries 
\begin{subequations}
\begin{align}
m^{\alpha, \beta}_{w} = \frac{2 \sum_i\sum_{j \neq i} \min \{ w_{ij}^{\alpha}, w_{ij}^{\beta}\}}{W^{\alpha} + W^{\beta}}=\frac{2 W^{\alpha\rightrightarrows\beta}}{W^{\alpha} + W^{\beta}}, & \\ 
r^{\alpha, \beta}_{w} = \frac{2 \sum_i\sum_{j \neq i} \min \{ w_{ij}^{\alpha}, w_{ji}^{\beta}\}}{W^{\alpha} + W^{\beta}}=\frac{2 W^{\alpha\rightleftarrows\beta}}{W^{\alpha} + W^{\beta}}, & 
\end{align}
\label{m_wei}
\end{subequations}
where $ W^{\alpha}=\sum_i\sum_{j \neq i}w_{ij}^\alpha $ is the total weight of the links in layer $ \alpha $ (analogously for layer $ \beta $), $ W^{\alpha\rightrightarrows\beta}=\sum_i\sum_{j \neq i} \min \{ w_{ij}^{\alpha}, w_{ij}^{\beta}\}$ is the total link weight of layer $\alpha$ that is multiplexed in layer $\beta$ (clearly, $ W^{\alpha\rightrightarrows\beta}= W^{\beta\rightrightarrows\alpha}$), and $ W^{\alpha\rightleftarrows\beta}=\sum_i\sum_{j \neq i} \min \{ w_{ij}^{\alpha}, w_{ji}^{\beta}\}$ is the total link weight of layer $\alpha$ that is reciprocated in layer $\beta$ (clearly, $ W^{\alpha\rightleftarrows\beta}= W^{\beta\rightleftarrows\alpha}$).  
The matrices ${\bf M}_w$ and ${\bf R}_w$ represent the two generalizations, for directed multiplexes, of the weighted multiplexity matrix introduced in~\cite{gemmetto} for undirected weighted multiplexes.
Like their binary counterparts, both matrices have entries in the range $ [0,1] $, the maximum value being attained by identical ($w_{ij}^\alpha=w_{ij}^\beta$ for all $i,j$) and fully `multireciprocated' ($w_{ij}^\alpha=w_{ji}^\beta$ for all $i,j$) layers respectively.
In analogy with the corresponding binary case, the diagonal of ${\bf M}_w$ has all unit entries while that of ${\bf R}_w$ has entries that coincide with the recent definition of reciprocity for weighted monoplex networks~\cite{Squartini}.\\

In this case as well, for trivial multiplexes with sparse noninteracting layers and narrow strength distributions, the two matrices are expected to be asymptotically diagonal. However, this is no longer true in presence of dense layers and/or for broad strength distributions, and we therefore need a comparison of the raw quantities with their expected value under a null model (now the DWCM). This consideration leads us to introduce the transformed weighted multiplexity and multireciprocity matrices with entries
\begin{subequations}
\begin{align}
\mu^{\alpha, \beta}_w &= \frac{m^{\alpha, \beta}_{w} - \langle m^{\alpha, \beta}_{w} \rangle_{\rm DWCM}}{1 - \langle m^{\alpha, \beta}_{w} \rangle_{\rm DWCM}},  \\ 
\rho^{\alpha, \beta}_w &= \frac{r^{\alpha, \beta}_{w} - \langle r^{\alpha, \beta}_{w} \rangle_{\rm DWCM}}{1 - \langle r^{\alpha, \beta}_{w} \rangle_{\rm DWCM}}, 
\end{align}
\label{mu_wei}
\end{subequations}
\noindent where $ \langle\cdot \rangle_{\rm DWCM}$ denotes the expected value under the DWCM. 
As in the binary case, we can derive an analytical expression for the expected values that ultimately requires the expectation of the minimum of $w_{ij}^\alpha$ and $w_{ij}^\beta$ (or $w_{ji}^\beta$). 
This is done in Appendix~\ref{app:DWCM}.
It turns out that, under the DWCM, the distribution of link weights is geometrical~\cite{Squartini,Squartini6}:
\begin{eqnarray}
P(w_{ij}^{\alpha}) = (p_{ij}^{\alpha})^{w_{ij}^{\alpha}} (1 - p_{ij}^{\alpha}),
\label{eq:Geometric}
\end{eqnarray}
where $p_{ij}^{\alpha}$ denotes again the probability that a directed link (of any positive weight) from node $i$ to node $j$ is realized in layer $\alpha$.
The above probability can be used to calculate $ \mu^{\alpha, \beta}_w$ and $ \rho^{\alpha, \beta}_w$ analytically as discussed in Appendix~\ref{app:DWCM}.\\

The weighted multireciprocity of the multiplex can be conveniently compared with the weighted reciprocity of the aggregated monoplex network. The link weights of the latter are defined by
\begin{equation}
w_{ij}^{\rm mono} = \sum_{\alpha = 1}^{M} w_{ij}^{\alpha},
\label{mono_w}
\end{equation} 
and the associated aggregate weighted reciprocity~\cite{Squartini} is
\begin{eqnarray}
r^{\rm mono}_{w} = \frac{\sum_i\sum_{j \neq i} \min \{ w_{ij}^{\rm mono}, w_{ji}^{\rm mono}\}}{W^{\rm mono}}
\label{r_wei_aggr}
\end{eqnarray}
(where $W^{\rm mono}=\sum_i\sum_{j \neq i}w_{ij}^{\rm mono}$). The corresponding filtered value $\rho^{\rm mono}_w$ can be defined as in~(\ref{mu_wei}b).\\

In analogy with the binary case, it is possible to define the $z$-scores associated to $ m^{\alpha, \beta}_{w} $ and $ r^{\alpha, \beta}_{w} $ as follows:
\begin{subequations}
\begin{align}
z \big( m^{\alpha, \beta}_{w} \big) = \frac{m^{\alpha, \beta}_{w} - \langle m^{\alpha, \beta}_{w} \rangle_{\rm DWCM}}{\sqrt{\langle ( m^{\alpha, \beta}_{w} )^2 \rangle_{\rm DWCM} - \langle m^{\alpha, \beta}_{w} \rangle_{\rm DWCM}^2}}, & \\ 
z \big( r^{\alpha, \beta}_{w} \big) = \frac{r^{\alpha, \beta}_{w} - \langle r^{\alpha, \beta}_{w} \rangle_{\rm DWCM}}{\sqrt{\langle ( r^{\alpha, \beta}_{w} )^2 \rangle_{\rm DWCM} - \langle r^{\alpha, \beta}_{w} \rangle_{\rm DWCM}^2}}. &
\end{align}
\label{zScores_wei}
\end{subequations}
The explicit analytical expressions for these $z$-scores are calculated in Appendix~\ref{app:DWCM}. 
Again, the $z$-scores \eqref{zScores_wei} have the same signs as the corresponding quantities \eqref{mu_wei}, but in addition they allow to test for statistical significance using e.g. a threshold of $z_c=2$.\\


\section{Empirical analysis of the World Trade Multiplex\label{sec:WTM}}
In this section, we apply the framework defined so far to the analysis of a real-world system. This system is the World Trade Multiplex (WTM), defined as the multi-layer network representing the directed trade relations between world countries in different commodities.
At both the binary and the weighted level, the structure of the aggregate (monoplex) version of this network is well studied \cite{garlaschelliloffredo2004,squartiniPRE2011a,squartiniPRE2011b}, as well as that of many of its layers separately \cite{barigozzi,mastrandreaPRE2014}. 
However, much less is known about the inter-layer dependencies in the WTM.
In particular, an assessment of the inter-layer couplings that are not simply explained by the local topological properties of the WTM has been carried out only for the undirected version of the network~\cite{gemmetto}.
Given the importance of the directionality of trade flows, especially at the disaggregated level of individual commodities, it is therefore important to carry out a directed analysis of the WTM.
The tools we have introduced in the previous section allow us to make this step and arrive at a novel characterization of the WTM where the undirected multiplexity properties documented in~\cite{gemmetto} are resolved into their two directed components, namely multiplexity and multireciprocity.
These results have important potential implications for problems related to research on international trade, such as the definition of trade-based `product taxonomies'~\cite{barigozzi}, the construction of the `product space'~\cite{Hidalgo}, and the calculation of `fitness and complexity' metrics~\cite{Tacchella}. These points are discussed later in sec.~\ref{sec:conclusions}.

\subsection{Data}
We use the BACI-Comtrade dataset~\cite{Baci} where international trade flows among all countries of the world are disaggregated into different commodity classes at the 2-digit resolution level, defined as in the standard HS1996 classification~\cite{Comtrade} of traded goods. 
It is possible to represent this dataset as a multiplex as in~\cite{barigozzi,gemmetto,mastrandreaPRE2014}.
In particular, we will consider a multi-layer representation defined by $N=207$ nodes (countries) and $M=96$ layers (commodities), for the year 2011.
Since each trade exchange is reported by both the importer and the exporter (and the two values may in general differ), the dataset uses a reconciliation procedure to get a unique value for each flow (see~\cite{Baci} for details). All the resulting trade volumes are expressed in thousands of dollars in the dataset. Since our approach works for integer link weights, all the reported trade values have been rescaled by first dividing by 10 and then rounding to the closest integer. This defines our integer link weights $\{w_{ij}^\alpha\}$ for all layers. For each entry $w_{ij}^\alpha$, we then define $a_{ij}^\alpha=1$ if $w_{ij}^\alpha>0$ and $a_{ij}^\alpha=0$ otherwise. We point out that the rounding procedure does not significantly affect the structure of the system under study, as the percentage of original links which are lost (i.e. rounded to zero) is negligible.\\

From the multiplex trade flows we also compute the aggregate binary and weighted links $a_{ij}^{\rm mono}$ and $w_{ij}^{\rm mono}$ between any two countries $ i $ and $ j $ in the collapsed monoplex trade network, as in \eqref{mono_b} and \eqref{mono_w} respectively. This allows us to compare the multiplex structure of trade with the aggregate one and highlight relevant information that is lost in the aggregation procedure.
For instance, for both the binary and the weighted representation of the system, we can compare the values of the multireciprocity matrix measured on the commodity-resolved multiplex with the usual scalar reciprocity measured on the monoplex aggregate trade network.

\subsection{Binary analysis}
\begin{figure*}[t]
\begin{center}
\includegraphics[width=1.0\textwidth]{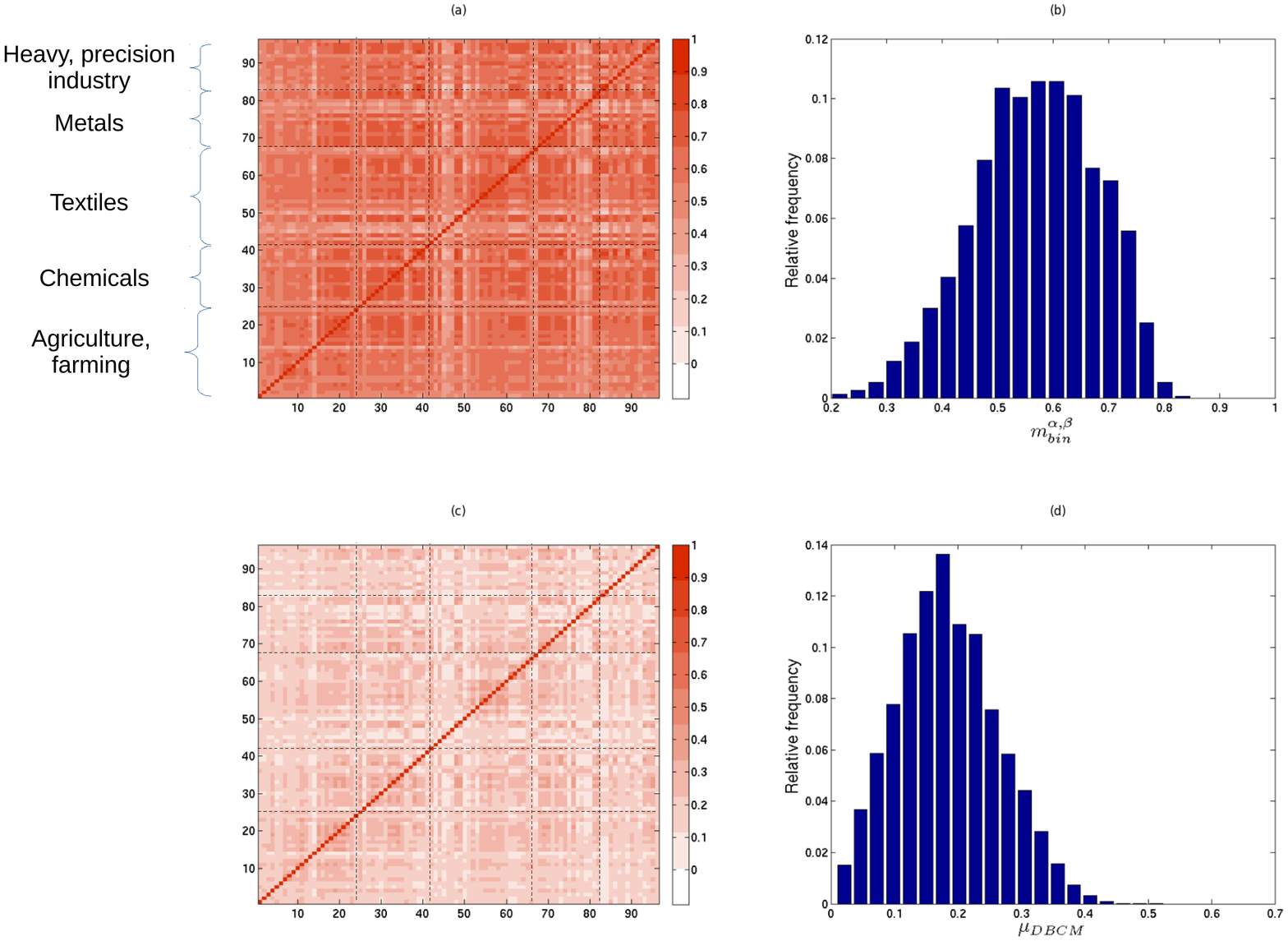}
\end{center}
\caption{(Color online) Top panels: color-coded binary multiplexity matrix ${\bf M}_b$ (a) and corresponding distribution of off-diagonal multiplexity values $ m^{\alpha, \beta}_b $ (with $\alpha\ne\beta$) (b). 
Bottom panels: 
same as for the top panels, but with raw binary multiplexity $ m^{\alpha, \beta}_b $ replaced by rescaled binary multiplexity $ \mu^{\alpha, \beta}_b $.}
\label{fig:m_bin_BACI}
\end{figure*}

We start with a binary analysis of the WTM, thus taking into account only the topology of the various layers while disregarding the information about trade volumes.  
In Figure~\ref{fig:m_bin_BACI}(a) we show the color-coded binary multiplexity matrix ${\bf M}_b$. Next to it, in Figure~\ref{fig:m_bin_BACI}(b) we show the corresponding frequency distribution of off-diagonal matrix entries $m_b^{\alpha,\beta}$ (with $\alpha\ne\beta$). In calculating the frequencies, we discard the diagonal entries because they trivially evaluate to $m_b^{\alpha,\alpha}=1$, as discussed above. 
High values of multiplexity are observed for most of the pairs of commodities.
This result is in agreement with what has been reported in~\cite{gemmetto} on the basis of an undirected analysis of the WTM where imports and exports between any two countries were combined together into a single trade link.\\

As we mentioned, the multiplexity matrix ${\bf M}_b$ would be asymptotically diagonal for trivial multiplexes with sparse non-interacting layers and narrow degree distributions. However, since the layers of the WTM are very dense and their degree distributions significantly broad \cite{barigozzi,gemmetto,mastrandreaPRE2014}, this system is an ideal case study requiring the use of a null model in order to assess the presence of a genuine coupling among layers.
In Figure~\ref{fig:m_bin_BACI}(c) we show the color-coded matrix of rescaled multiplexity values $\mu^{\alpha, \beta}_b$, which control for the effects of the heterogeneity of the layer-specific in- and out-degree sequences. Similarly, in Figure~\ref{fig:m_bin_BACI}(d) we show the corresponding distribution of off-diagonal entries. 
We find that, after controlling for the degrees, a significant amount of correlation is destroyed. However all the values are still strictly positive, indicating a tendency of all pairs of commodities to be `traded together'. The statistical significance of this result is discussed later in terms of $z$-scores.\\

\begin{figure*}[htbp]
\begin{center}
\includegraphics[width=1.0\textwidth]{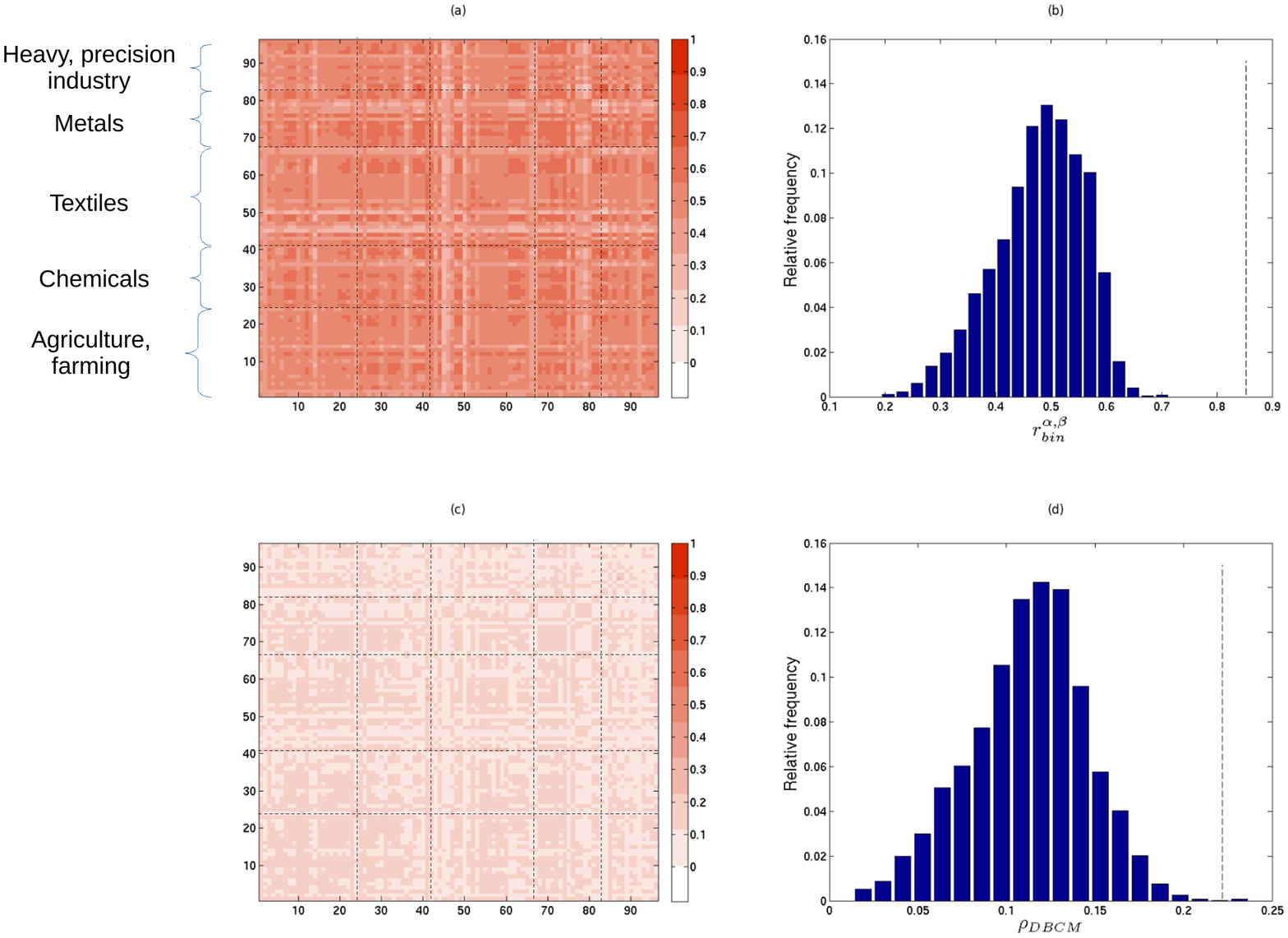}
\end{center}
\caption{(Color online) Top panels: color-coded binary multireciprocity matrix ${\bf R}_b$ (a) and corresponding distribution of off-diagonal multireciprocity values $ r^{\alpha, \beta}_b $ (with $\alpha\ne\beta$) (b). 
Bottom panels: 
same as for the top panels, but with raw binary multireciprocity $ r^{\alpha, \beta}_b $ replaced by rescaled binary multireciprocity $ \rho^{\alpha, \beta}_b $. The dashed lines represent the value of (raw and rescaled) binary reciprocity $r_b^{\rm mono}$ and $\rho_b^{\rm mono}$ of the aggregated monoplex network.}
\label{fig:r_bin_BACI}
\end{figure*}

We now move to the analysis of multireciprocity.
It is known that, when the aggregate trade in all commodities is considered, the binary monoplex representation of the World Trade Network exhibits a high level of reciprocity~\cite{Garlaschelli2,ruzzenentiSYMMETRY2,piccioloIEEE2012}.
It is interesting to see whether such property is preserved also at the multiplex level, and how the values compare with the aggregate case. 
Figure~\ref{fig:r_bin_BACI}(a) shows the color-coded binary multireciprocity matrix ${\bf R}_b$ and Figure~\ref{fig:r_bin_BACI}(b) the corresponding distribution of off-diagonal entries \footnote{We discard the diagonal entries in order to make the distribution compatible with the corresponding distribution for the multiplexity shown above; in any case, if the diagonal entries are included, the distribution looks very similar.}, with a superimposed delta function indicating the value of the binary reciprocity $r^\textrm{mono}_b$ of the aggregate monoplex network as a comparison.
The results are comparable with those found above for the multiplexity. 
Also in this case, the high multireciprocity values are consistent with the high multiplexity values found for the undirected representation of the WTM \cite{gemmetto} (where pairs of reciprocated links in each layer are merged into single undirected links).
However, for the multireciprocity this result is much less trivial than for the multiplexity, given the chosen level of disaggregation into many commodity classes. 
Indeed one would expect that, at such a relatively high resolution, it should be not very likely (at least not as likely as in the undirected representation) that the same commodity is traded ``back and forth'', i.e. both ways between the same two countries.
In any case we do find, in accordance with what we expect, that for all pairs of commodities the multireciprocity is significantly smaller than the reciprocity  $r^\textrm{mono}_b$ of the aggregate monoplex. This means that, as layers are aggregated, there is a bigger relative increment (with respect to individual layers) in the overall number of reciprocated links than in the total number of links. \\

As an interesting result, the intra-layer reciprocity values  $r^{\alpha,\alpha}_b$ lying along the diagonal of the multireciprocity matrix are found to be very similar to the values of the matrix entries  $r^{\alpha,\beta}_b$ lying close to the diagonal. Indeed, in the matrix plot of Figure~\ref{fig:r_bin_BACI}a the diagonal is visually indistinguishable from the entries of the matrix that are ``nearby''. 
Given the order of the commodities in the matrix (as shown for instance in~\cite{gemmetto}), these nearby entries represent the multireciprocity between pairs of similar commodities.
This result means that the high reciprocity of the aggregate trade monoplex does \emph{not} arise from the superposition of layers with high internal reciprocity and low mutual multireciprocity (as would be the case in presence of an approximately diagonal multireciprocity matrix). 
Rather, we find that a trade flow in one commodity $\alpha$ tends to be reciprocated by comparable trade flows in several different commodities, including (but not dominated by) the same commodity $\alpha$ and many other related commodities.
Specifically, it can be seen from Figure~\ref{fig:r_bin_BACI}a  that layers characterized by low (high) values of internal reciprocity are embedded within groups of layers with low (high) mutual multireciprocity. 
This suggests that the level of reciprocity in international trade is not an intrinsic property of individual commodities, but rather a property of whole groups of mutually reciprocated commodities with comparable multireciprocity values.\\

In Figure~\ref{fig:r_bin_BACI}(c) and (d) we show the color-coded binary rescaled multireciprocity matrix and the corresponding distribution of off-diagonal entries $ \rho^{\alpha, \beta}_b $ (with $\alpha\ne\beta$). 
The relatively small values (with respect to the non-rescaled quantities) indicate that, in analogy with what we found for the multiplexity, the apparent correlation between the topology of pairs of layers is largely encoded in the relatedness of the degree sequences of such pairs. 
For the vast majority of pairs of commodities the multireciprocity is still lower than that measured on the aggregate network.
However, all pairs of layers preserve a positive residual multireciprocity, the statistical significance of which is studied later in our $z$-score analysis.\\

\begin{figure*}[htbp]
\begin{center}
\includegraphics[width=1.0\textwidth]{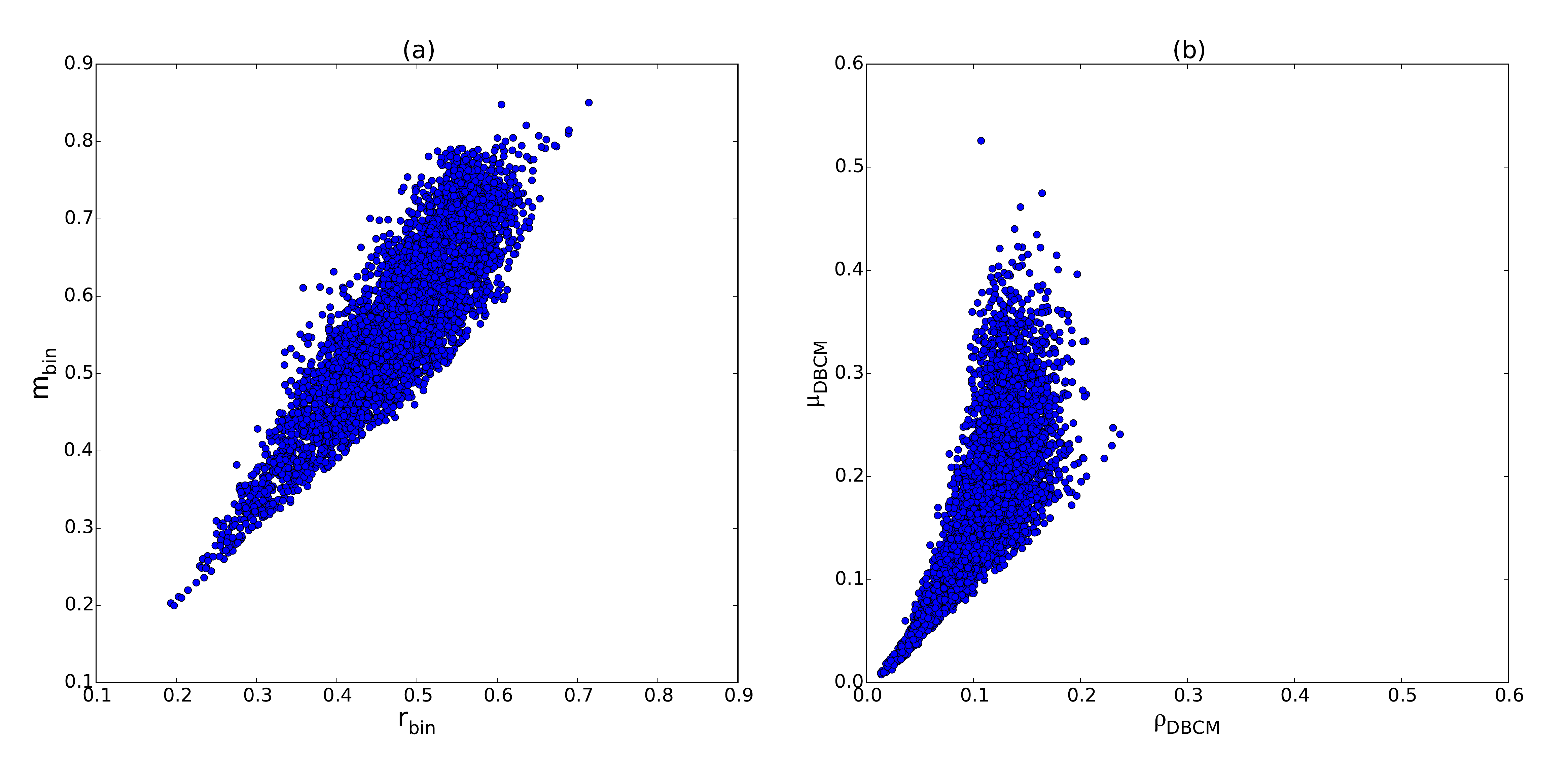}
\end{center}
\caption{(Color online) Scatter plots of off-diagonal binary multireciprocity values versus off-diagonal binary directed multiplexity values. Left: raw values ($ r^{\alpha, \beta}_b $ vs $ m^{\alpha, \beta}_b $); 
right: rescaled values ($ \rho^{\alpha, \beta}_b $ vs $ \mu^{\alpha, \beta}_b $).}
\label{fig:rec_mul_bin_BACI}
\end{figure*}

When we look at the multiplexity matrix in Figure~\ref{fig:m_bin_BACI}(a) and the corresponding multireciprocity matrix in Figure~\ref{fig:r_bin_BACI}(a), 
we see the appearance of similar patterns. Such similarity is further investigated in Figure~\ref{fig:rec_mul_bin_BACI}(a), where we 
report the scatter plots of pairwise multireciprocity values versus the corresponding multiplexity values. 
We observe a roughly linear trend, which is however lost when we look at the filtered values, as shown in Figure~\ref{fig:rec_mul_bin_BACI}(b). 
We see that, in the latter case, the relationship between $ \rho^{\alpha, \beta}_b $ and $ \mu^{\alpha, \beta}_b $ is non-linear and significantly scattered.
Although the presence of a non-linear relation may be related to the particular choice of normalization adopted in~(\ref{def_mu_bin}), we point out that the entity of the scatter is so big that it is not possible to retrieve the value of multiplexity from the multireciprocity, and vice-versa.
This illustrates that the two quantities convey different pieces of information that are irreducible to each other.\\ 

\begin{figure*}[htbp]
\begin{center}
\includegraphics[width=1.0\textwidth]{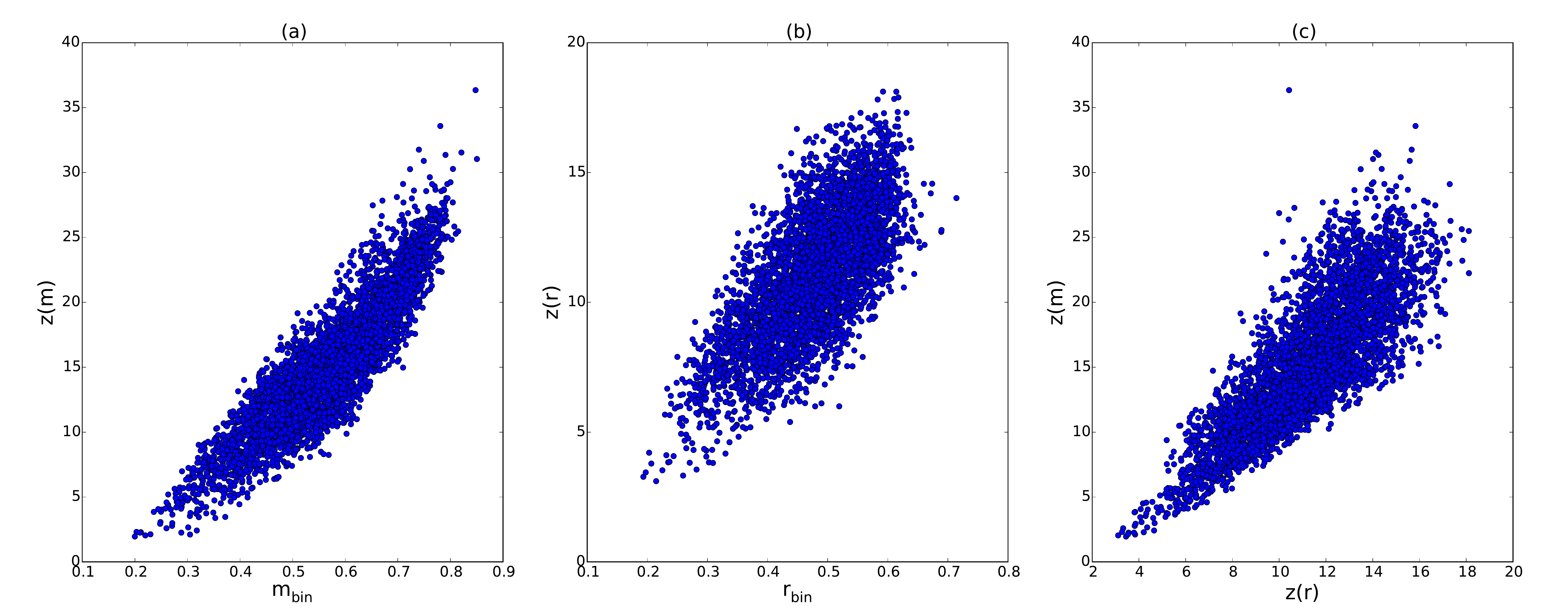}
\end{center}
\caption{(Color online) Left: binary transformed multiplexity $ \mu^{\alpha, \beta}_b $ versus its corresponding $z$-score $ z(m^{\alpha, \beta}_b) $; center: binary transformed multireciprocity $ \rho^{\alpha, \beta}_b $ versus its corresponding $z$-score $ z(r^{\alpha, \beta}_b) $; right: $ z(r^{\alpha, \beta}_b) $ vs $ z(m^{\alpha, \beta}_b) $. Only off-diagonal values are reported.}
\label{fig:zScores_BACI}
\end{figure*}

Similar considerations apply to the $z$-scores. In Figure~\ref{fig:zScores_BACI}(a) and~\ref{fig:zScores_BACI}(b) we show the empirical relation between the transformed multiplexity and multireciprocity and their corresponding $z$-scores: it is worth recalling that the information provided by these two quantities can be \emph{a priori} different, given the lack of information about the standard deviation in the rescaled multiplexity and multireciprocity metrics. 
Empirically, we however find a strong correlation between these quantities, indicating that large values of binary multiplexity or multireciprocity correspond to large $z$-scores, and vice-versa.
Moreover, even the smallest $z$-scores (those found for the
pairs of layers showing very low multiplexity or multireciprocity) are still quite high (i.e. positive and larger than $z_c=2$) in terms of statistical significance. This means that even the pairs of layers with smallest multiplexity or multireciprocity should be considered as significantly and positively correlated.
We therefore conclude that, at a binary level, every commodity of the WTM tends to be traded together with all other commodities, both in the same  and in the opposite direction. 
As we show below, this is no longer the case when the weighted version of the multiplex is considered.\\

Figure~\ref{fig:zScores_BACI}(c) shows the relation existing between $ z(r^{\alpha, \beta}_b) $ and $ z(m^{\alpha, \beta}_b) $ for each pair of layers. 
 If we compare this figure with Figure~\ref{fig:rec_mul_bin_BACI}, we see that in this case the trend is more linear, although the scatter is again quite large.
This confirms that it is not possible to recover the values of multiplexity from those of multireciprocity, and vice-versa.

\subsection{Weighted analysis}

We now perform a weighted analysis of the World Trade Multiplex, by taking into account the value of imports and exports observed between countries.\\

\begin{figure*}[htbp]
\begin{center}
\includegraphics[width=1.0\textwidth]{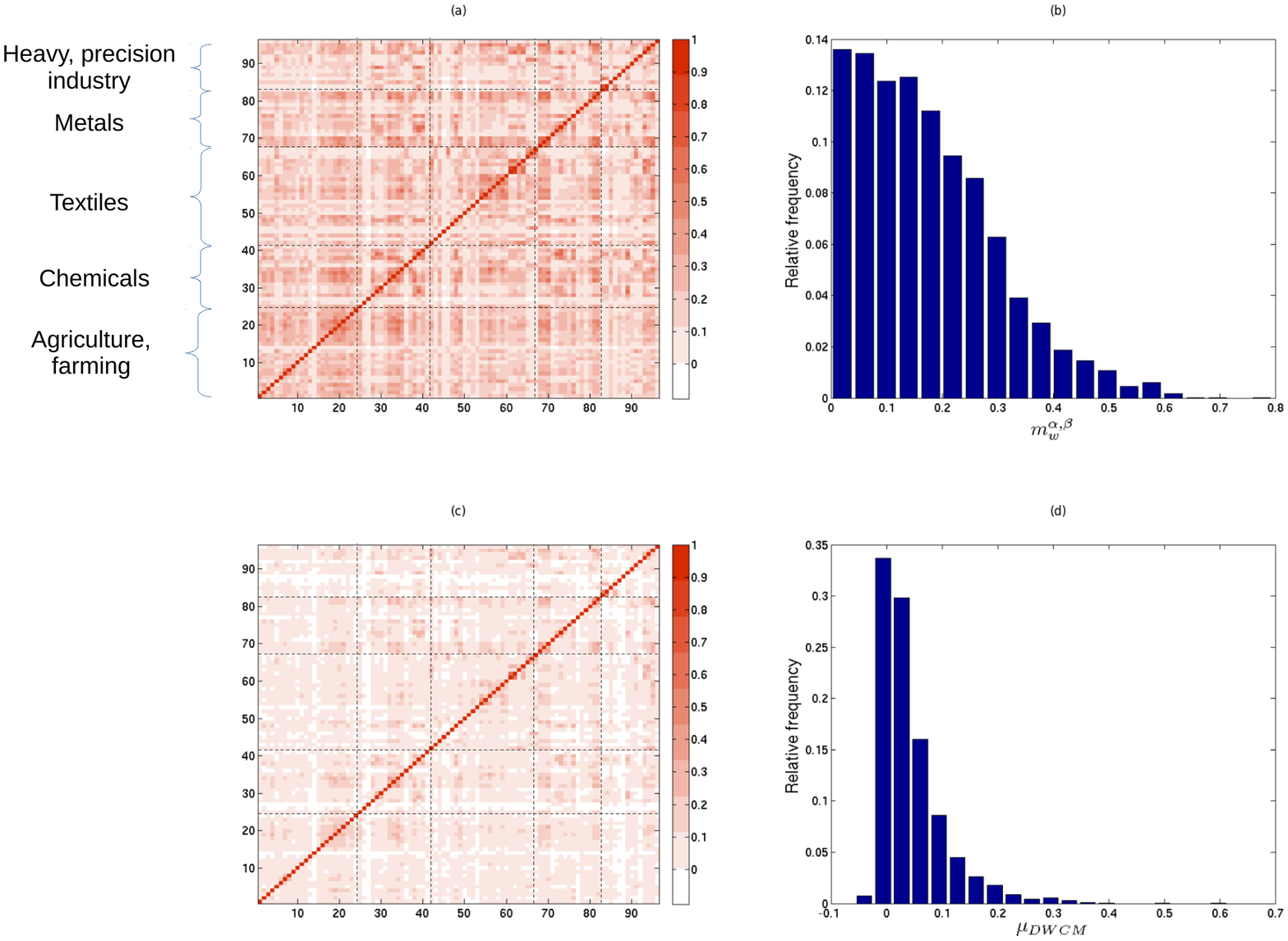}
\end{center}
\caption{(Color online) Top panels: color-coded weighted multiplexity matrix ${\bf M}_w$ (a) and corresponding distribution of off-diagonal multiplexity values $ m^{\alpha, \beta}_w $ (with $\alpha\ne\beta$) (b). 
Bottom panels: 
same as for the top panels, but with raw weighted multiplexity $ m^{\alpha, \beta}_w $ replaced by rescaled weighted multiplexity $ \mu^{\alpha, \beta}_w $. Note that, in panel (c), white entries represent negative values.}
\label{fig:m_wei_BACI}
\end{figure*}

In Figure~\ref{fig:m_wei_BACI}(a) and (b) we show the color-coded weighted directed multiplexity matrix ${\bf M}_w$ and the distribution of its off-diagonal entries. 
We clearly see that, even though several pairs of commodities are still strongly overlapping, the multiplexity distribution is concentrated over a range of significantly smaller values with respect to the corresponding binary distribution. 
Indeed, the notion of weighted multiplexity, by involving the minimum of the weights of two reciprocated links, provides a stricter criterion with respect to the unweighted case. 
In particular, for any pair of nodes and any pair of layers, it is more unlikely to achieve the maximum weighted value $\textrm{min}\{w_{ij}^\alpha,w_{ij}^\beta\}$ than the maximum binary value $\textrm{min}\{a_{ij}^\alpha,a_{ij}^\beta\}$.
Lower values of multiplexity with respect to Figure~\ref{fig:m_bin_BACI}(a) are therefore expected. We also expect to find a similar reduction for the multireciprocity later on.\\

In Figure~\ref{fig:m_wei_BACI}(c) and (d) we report the color-coded weighted rescaled multiplexity matrix and the corresponding distribution of off-diagonal entries 
$ \mu^{\alpha, \beta}_w $.
The fact that many values are now mapped to zero means that a significant component of the overlap between commodities can  be explained simply in terms of the correlated strength sequences of the various layers. 
Importantly, we see that some pairs of layers actually exhibit negative rescaled multiplexity, even though the distribution is far from symmetric.
This result, which is only visible in the weighted analysis, means that there are pairs of commodities for which the observed trade multiplicity is actually \emph{lower} than expected under the null model: these commodities prefer `not to be traded together'.\\

\begin{figure*}[htbp]
\begin{center}
\includegraphics[width=1.0\textwidth]{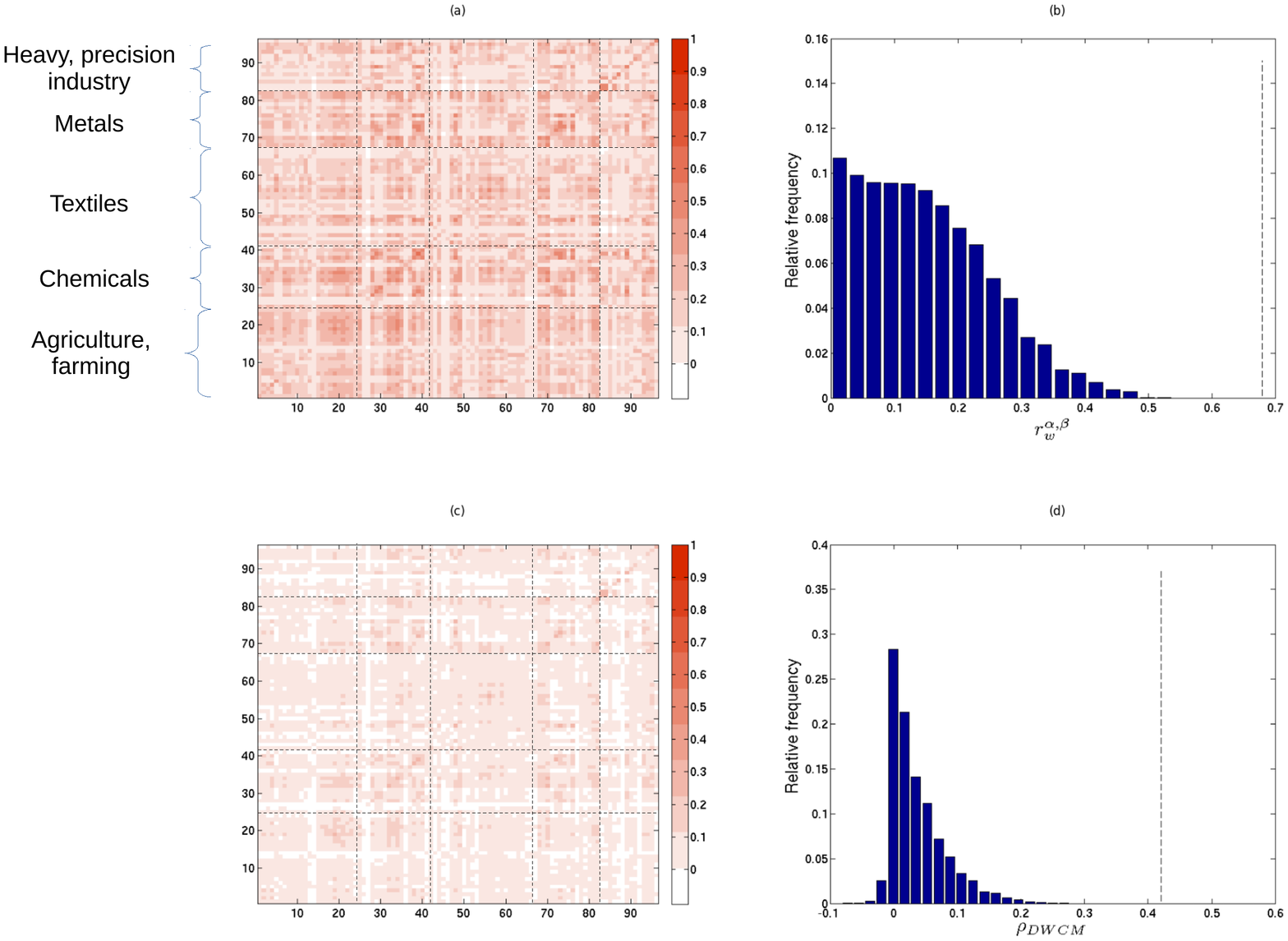}
\end{center}
\caption{(Color online) Top panels: color-coded weighted multireciprocity matrix ${\bf R}_w$ (a) and corresponding distribution of off-diagonal multireciprocity values $ r^{\alpha, \beta}_w $ (with $\alpha\ne\beta$) (b). 
Bottom panels: 
same as for the top panels, but with raw weighted multireciprocity $ r^{\alpha, \beta}_w $ replaced by rescaled weighted multireciprocity $ \rho^{\alpha, \beta}_w $. The dashed lines represent the value of (raw and rescaled) weighted reciprocity $r_w^{\rm mono}$ and $\rho_w^{\rm mono}$ of the aggregated monoplex network. Note that, in panel (c), white entries represent negative values.}
\label{fig:r_wei_BACI}
\end{figure*}

We then analyze the weighted multireciprocity of the WTM. 
Recently, it has been shown that the aggregated version of the network has a strong weighted reciprocity~\cite{Squartini}, a result that we can now complement with the analysis of the disaggregated multiplex. 
In Figure~\ref{fig:r_wei_BACI}(a) and (b) we report the color-coded weighted multireciprocity matrix ${\bf R}_w$, along with the distribution of its off-diagonal entries.
In analogy with the binary case, we see that the aggregated network exhibits a reciprocity which is significantly higher than the multireciprocity associated to any individual pair of layers.
Yet several pairs of commodities are characterized by a substantial level of multireciprocity.
In Figure~\ref{fig:r_wei_BACI}(c) and (d) we show the corresponding results for the rescaled weighted multireciprocity $\rho_w^{\alpha,\beta}$. We see that many values become close to zero and some become negative, in analogy with the behaviour of the multiplexity. The identification of pairs of layers with negative rescaled multireciprocity indicates that the corresponding commodities `prefer not to be traded in opposite directions', in contrast with the results we found in the binary analysis.\\

\begin{figure*}[htbp]
\begin{center}
\includegraphics[width=1.0\textwidth]{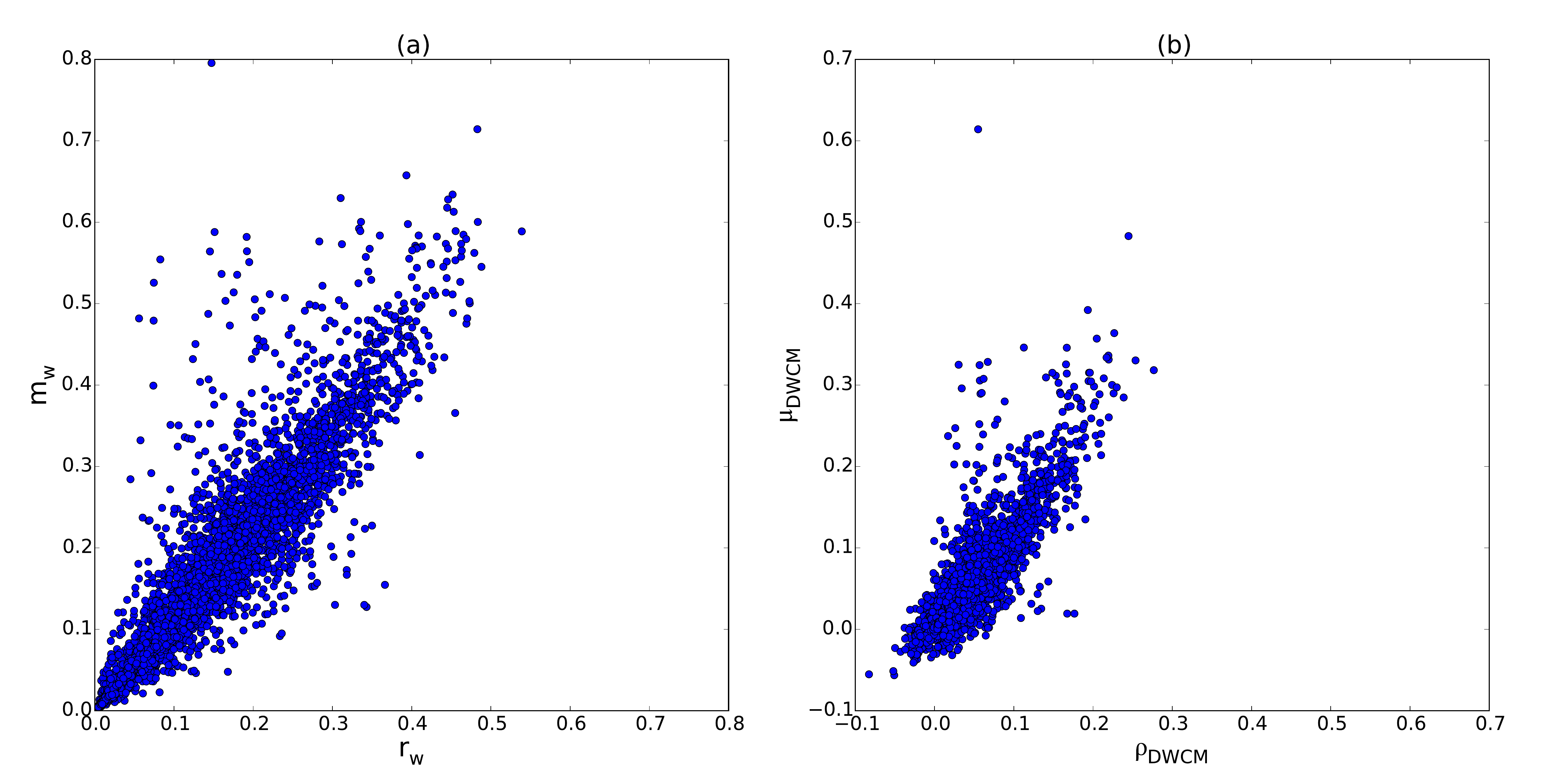}
\end{center}
\caption{(Color online) Scatter plots of off-diagonal weighted multireciprocity values versus off-diagonal weighted directed multiplexity values. Left: raw values ($ r^{\alpha, \beta}_{w} $ vs $ m^{\alpha, \beta}_{w} $); 
right: rescaled values ($ \rho^{\alpha, \beta}_w $ vs $ \mu^{\alpha, \beta}_w $).}
\label{fig:rec_mul_wei_BACI}
\end{figure*}

In Figure~\ref{fig:rec_mul_wei_BACI} we compare the weighted multireciprocity and the weighted multiplexity. When we consider the raw values (a), we observe a clear linear 
trend (although more scattered than in the corresponding unweighted case). The trend becomes even more robust, and less noisy, for the filtered values, as shown in (b).
In both panels, the most significant commodities (both in terms of trade volumes and economic relevance) mainly lie along the diagonal, while the outliers represent less relevant products (for instance, some textiles or less traded craft goods). 
We also see pairs of commodities whose multireciprocity is similar to the reciprocity of the aggregate trade network. These commodities, such as cereals and heavy industry products, are not necessarily the most traded ones, still they better represent the reciprocity patterns of total trade among countries, possibly because they give the main contribution to the reciprocity of the aggregated network.\\

\begin{figure*}[htbp]
\begin{center}
\includegraphics[width=1.0\textwidth]{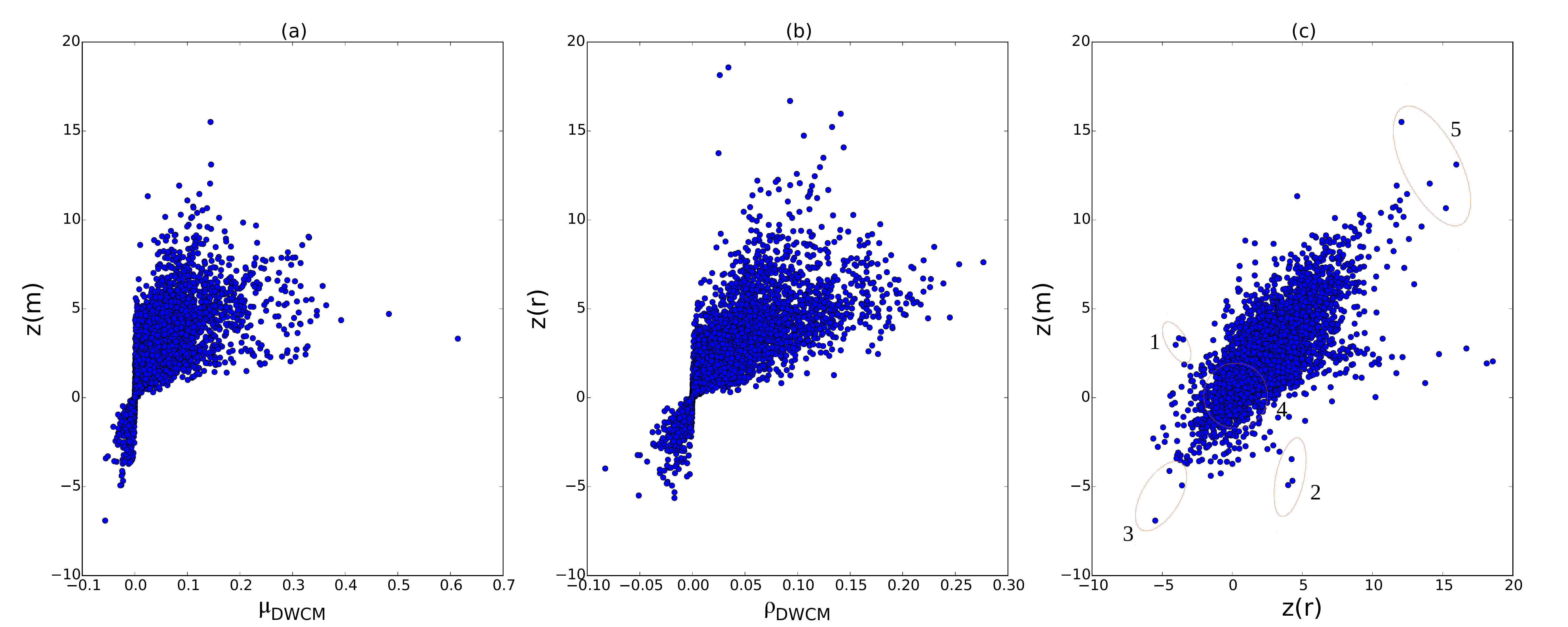}
\end{center}
\caption{(Color online) Left: weighted transformed multiplexity $ \mu^{\alpha, \beta}_w $ versus its corresponding $z$-score $ z(m^{\alpha, \beta}_{w}) $; center: weighted transformed multireciprocity $ \rho^{\alpha, \beta}_w $ versus its corresponding $z$-score $ z(r^{\alpha, \beta}_{w}) $; right: $ z(r^{\alpha, \beta}_{w}) $ vs $ z(m^{\alpha, \beta}_{w}) $. In panel (c), numbered circles correspond to the bullet points reported in Sec. III C. Only off-diagonal values are reported.}
\label{fig:zScores_BACI_wei}
\end{figure*}

Quantitatively, another important difference between the binary and the weighted approach lies in the statistical significance of the values of multiplexity and multireciprocity, as we can see from the analysis of the $z$-scores (Figure~\ref{fig:zScores_BACI_wei}). Indeed, in the unweighted case we found that even the smallest values of $ \mu^{\alpha, \beta}_b $ and $ \rho^{\alpha, \beta}_b $ are significant, as the corresponding $z$-scores are larger than the critical value $z_c$. 
Instead, here we observe almost no correlation (except for the aforementioned sign concordance) between weighted multiplexity or multireciprocity and the corresponding $z$-scores (see Fig.~\ref{fig:zScores_BACI_wei}(a) and \ref{fig:zScores_BACI_wei}(b) respectively). 
Indeed, the same value of $ \mu^{\alpha, \beta}_b $ or $ \rho^{\alpha, \beta}_b $ may even correspond to $z$-scores with  different orders of magnitude. 
This means that, even for two pairs of layers with the same observed value of weighted multiplexity or multireciprocity, the statistical significance of the inter-layer coupling can be very different.
Moreover, the absolute value of many weighted $z$-scores is found below the significance threshold $z_{c}= 2 $, identifying pairs of uncorrelated layers (a result that is unobserved in the binary case). 
Finally, many pairs of commodities have a negative $z$-score below $-z_c$ for the multiplexity and/or multireciprocity. For these pairs, the tendency \emph{not} to be traded in the same direction and/or in opposite direction is statistically validated and confirms a difference with respect to the binary case.\\

As a final result, in Figure~\ref{fig:zScores_BACI_wei}(c) we show the relation existing between $ z \big( m^{\alpha, \beta}_{w} \big) $ and $ z \big( r^{\alpha, \beta}_{w} \big) $. We find an overall level of correlation which however leaves room for a significant scatter of points around the identity line. This scatter is big enough to imply that, for a given significance threshold $z_c$, the pairs of commodities can be partitioned in the following five classes: 
\begin{enumerate}
\item a few pairs of commodities that tend to be traded in the same direction ($z(m_w^{\alpha,\beta})>z_c$) but not in opposite directions ($z(r_w^{\alpha,\beta})<-z_c$): examples are apparel articles vs ships and boats; food industry residues, prepared animal feed vs ores, slag and ash;
\item a few pairs of commodities that tend to be traded in opposite directions ($z(r_w^{\alpha,\beta})>z_c$) but not in the same direction ($z(m_w^{\alpha,\beta})<-z_c$): examples are ores, slag and ash vs footwear and gaiters; apparel articles vs ores, slag and ash;
\item a moderately-sized group of pairs of commodities that tend to be traded neither in the same direction ($z(m_w^{\alpha,\beta})<-z_c$) nor in opposite ones ($z(r_w^{\alpha,\beta})<-z_c$): examples are raw hides and skins vs arms and ammunitions; tobacco vs ships and boats;
\item a large group of pairs of commodities for which there is no statistically significant tendency in at least one of the two directions ($|z(m_w^{\alpha,\beta})|<z_c$ and/or 
$|z(r_w^{\alpha,\beta})|<z_c$): examples are tobacco vs inorganic chemicals; explosives, pyrotechnic products vs vehicles (note that this class can be further split in sub-classes where commodities are uncorrelated in one direction but correlated in different ways in the other direction);
\item a very large group of pairs of commodities that tend to be traded both in the same direction ($z(m_w^{\alpha,\beta})>z_c$) and in opposite ones ($z(r_w^{\alpha,\beta})>z_c$): examples are sugar vs cocoa; soap, waxes, candles vs sugar.
\end{enumerate}
It should be noted that, in contrast with the above classification, the binary analysis concluded that all pairs of commodities belong to the last class only.\\

\section{Discussion and conclusions\label{sec:conclusions}}
The study of multi-layer networks has received substantial attention in the last few years, leading to the introduction of several novel quantities characterizing the structure of multiplexes as well as the behaviour of several dynamical processes taking place on them. 
The aim of all these studies is that of highlighting the role of the inter-layer couplings, the latter being the ultimate reason why layers of a multiplex should be analysed together in the first place, rather than separately.
In this paper we have argued that even the \emph{simplest} definitions of inter-layer coupling, based merely on the structural overlap of links across layers, are strongly biased by the density, finiteness, and heterogeneity of the network.
We have shown that controlling for the above effects requires a quite elaborate statistical treatment.
Focusing on multiplexes with (binary or weighted) directed links, we have introduced maximum-entropy multiplex ensembles with given node properties as the unbiased null models serving as a benchmark for the empirically observed properties.
We have then defined novel multiplexity and multireciprocity metrics, respectively quantifying the tendency of pairs of links to `align' and/or `anti-align' across each pair of layers of a real-world directed multiplex.
Since links can exist in both directions in every layer, the possible tendencies of forming aligned (multiplexed) and anti-aligned (multireciprocated) links do not conflict with each other and can actually coexist. 
Both multiplexity and multireciprocity are matrix-valued, as they represent the possible couplings among all pairs of layers. While multiplexity is a natural extension of the corresponding definition for undirected multiplexes, multireciprocity is a novel concept representing a nontrivial extension of the notion of single-layer reciprocity to multi-layer networks.\\

We believe that our results can be of value for several applications.
For instance, they provide a statistically rigorous way to identify possible (groups of) layers that are uncorrelated from the other layers, thus allowing to simplify the whole multiplex into mutually independent sub-systems with smaller numbers of layers. This problem has received significant attention recently \cite{Dedomenico, Iacovacci}. Our finding of a strong influence of the local node properties on the overall level of inter-layer coupling suggests that many of the results found with alternative techniques that do not control for these effects might be subject to an uncontrolled level of bias.\\

Other more specific applications are relevant for the specific case study of the WTM.
In extreme summary, our detailed analysis of this system confirmed that its multiplex structure contains much more information than the aggregated network of total trade does. 
At a binary level, we found that all pairs of commodities tend to be traded together between countries, both in the same direction (high multiplexity) and in opposite directions (high multireciprocity).
At a weighted level, this result only holds for a subset of pairs of commodities. Other commodity pairs are not correlated and others even tend to avoid being traded together in the same direction and/or in opposite ones.
The multireciprocity structure of the WTM highlights a tendency of groups of commodities to have a comparably high mutual reciprocity, of the same entity of the internal single-layer reciprocity of these commodities.
When aggregated into the monoplex network of total international trade, the WTM has a resulting reciprocity that is much bigger than the multireciprocity among its constituent layers.\\

In the light of the above results, our approach has implications relevant to various directions in international trade research. In particular, it indicates concrete ways to refine existing measures of inter-commodity correlation or similarity that are widely used to construct, among others, `product taxonomies'~\cite{barigozzi}, the `product space'~\cite{Hidalgo} and `fitness and complexity' metrics~\cite{Tacchella}.
All these applications are briefly explained below.\\

Inter-commodity correlation metrics have been introduced to quantify the coupling among layers of the WTM~\cite{barigozzi}, with the goal of constructing `product taxonomies' that reflect empirical trade similarities, as opposed to pre-defined product categories. 
However, as already pointed out in~\cite{gemmetto}, correlation metrics make an implicit and totally unrealistic assumption of structural homogeneity of the network, by interpreting all the edges of a layer as independent observations drawn from the same probability distribution. 
Our results provide alternative metrics of inter-layer coupling that replace the  homogeneity assumption with a much more realistic null model that accurately controls for the observed degree of node heterogeneity in each layer.
The use of our metrics is likely to change the structure of correlation-based product taxonomies significantly.\\

The `product space' is defined as a network of commodities connected by links whose weight quantifies the tendency of a pair of commodities to be traded together (in the same direction) between the same two countries~\cite{Hidalgo}. Our results clearly indicate that, to be statistically reliable, such an analysis should include a way to filter out the strong empirical heterogeneity of node degrees and/or node strengths.
Moreover, they highlight a second layer of information that should be relevant for the product space construction, namely the fact that, besides the tendency of pairs of commodities to be traded together in the same direction (multiplexity), there can be a substantial tendency of being traded in the opposite direction (multireciprocity). 
We found that these two effects have a comparable magnitude. We also found that pairs of commodities with approximately the same multiplexity can be characterized by very different levels of multireciprocity. 
This suggests that neglecting multireciprocity in the construction of the product space can represent a substantial loss of information.\\

Finally, the `fitness and complexity' approach focuses on the bipartite network of countries and their exported products, and uses the structure of this network to recursively define metrics of product complexity and country competitiveness (fitness)~\cite{Tacchella}. This method can reveal the `hidden' potential of countries that is not (yet) reflected in their current GDP levels.
Clearly, the output of this approach entirely depends on how the bipartite country-product matrix is constructed. 
This matrix is ultimately a projection of the WTM but is generally filtered using a null model based on the concept of `revealed comparative advantage' \cite{Balassa}, which however operates at the aggregate country-product level and not at the level of the underlying multiplex. As such, it does not control for the size of importers.
Our approach provides a way to enforce a more accurate null model on the original WTM and obtain an alternative bipartite country-product projection.\\

We believe that all the research directions outlined above deserve future explorations and we expect the results reported int this paper to be of use.

\begin{acknowledgments}
This work was supported by the EU project MULTIPLEX (contract 317532).
DG acknowledges support from the Netherlands Organization for
Scientific Research (NWO/OCW) and the Dutch Econophysics Foundation
(Stichting Econophysics, Leiden, the Netherlands) with funds from
beneficiaries of Duyfken Trading Knowledge BV, Amsterdam, the
Netherlands. TS is supported by the FET project DOLFINS (Grant No 640772).
\end{acknowledgments}

\appendix

\section{Maximum Entropy Method}
\label{app:maxent}
We define null models of multiplexes as canonical maximum-entropy ensembles satisfying a given set $\overrightarrow{\mathcal{C}}$ of $\mathcal{K}$ constraints on average. 
If $G^\alpha\in \mathcal{G}_N$ denotes the graph realized in layer $\alpha$ of the multiplex (recall that $\mathcal{G}_N$ is the set of all directed monoplex graphs with $N$ nodes), and if $\overrightarrow{G}\in \mathcal{G}_N^M$ denotes the entire multiplex (where $\mathcal{G}_N^M$ is the set of all directed multiplex graphs with $N$ nodes and $M$ layers), we write $\overrightarrow{G}=(G^\alpha)_{\alpha=1}^M$.
Now let $\overrightarrow{\mathcal{C}}$ denote a vector-valued function on $\mathcal{G}_N^M$, evaluating to $\overrightarrow{\mathcal{C}}(\overrightarrow{G})$ on the particular multiplex $\overrightarrow{G}$.
The vector $\overrightarrow{\mathcal{C}}(\overrightarrow{G})$ is to be regarded as a set of structural properties measured on $\overrightarrow{G}$.\\

A canonical ensemble of binary (weighted) directed multiplex networks with the \emph{soft constraint} $\overrightarrow{\mathcal{C}}$ is specified by a probability distribution $\mathcal{P}\big(\overrightarrow{G}|\overrightarrow{\theta}\big)$ on $\mathcal{G}_{N}^{M}$, where $\overrightarrow{\theta}$ is a vector of Lagrange multipliers required to enforce a desired expected value
\begin{equation}
\langle\overrightarrow{\mathcal{C}}\rangle_{\overrightarrow{\theta}}=
\sum_{\overrightarrow{G}\in\mathcal{G}_N^M}
\mathcal{P}\big(\overrightarrow{G}|\overrightarrow{\theta}\big)\overrightarrow{\mathcal{C}}(\overrightarrow{G})
\label{eq:constraintmulti}
\end{equation}
of $\overrightarrow{\mathcal{C}}$. 
Note that both $\overrightarrow{\theta}$ and  $\overrightarrow{\mathcal{C}}$ are vectors of numbers with the same (but model-dependent) dimension $\mathcal{K}$, while $\overrightarrow{G}$ is always an $M$-dimensional vector of graphs. 
Obviously, an additional constraint on the probability is the normalization condition
\begin{equation}
\sum_{\overrightarrow{G}\in\mathcal{G}_N^M}\mathcal{P} \big(\overrightarrow{G} |\overrightarrow{\theta}\big) =1\qquad \forall~\overrightarrow{\theta}.
\label{eq:normjoint}
\end{equation}\\

We want our ensembles to produce multiplexes with independent layers, as defined in Eq.~\eqref{prob_multiplex}.
This requirement corresponds to the enforcement of separate constraints on the different layers, i.e. $\overrightarrow{\mathcal{C}}=(\overrightarrow{{C}^\alpha})_{\alpha=1}^M$, where $\overrightarrow{{C}^\alpha}$ is a $K^\alpha$-dimensional vector of structural properties of the network in layer $\alpha$ only, evaluating to $\overrightarrow{{C}^\alpha}(G^\alpha)$ on the particular single-layer graph $G^\alpha$.
This leads to a separation in the corresponding Lagrange multipliers, i.e. $\overrightarrow{\theta}=(\overrightarrow{\theta^\alpha})_{\alpha=1}^M$.
$K^\alpha$ is the dimension of both $\overrightarrow{{C}^\alpha}$ and $\overrightarrow{\theta^\alpha}$, and we must have $\sum_{\alpha=1}^MK^\alpha=\mathcal{K}$.
Consequently, we can express the entropy of the ensemble of multiplex networks as
\begin{eqnarray}
 \mathcal{S} \big(\overrightarrow{\theta} \big) &\equiv& -\sum_{\overrightarrow{G}\in\mathcal{G}_{N}^{M}}  \mathcal{P}\big(\overrightarrow{G} | \overrightarrow{\theta}\big) \ln  \mathcal{P}\big(\overrightarrow{G} | \overrightarrow{\theta}\big)\nonumber\\
&=&\sum_{\alpha=1}^M S^\alpha\big(\overrightarrow{\theta^\alpha}\big),
\label{entropy}
\end{eqnarray}
where  
\begin{eqnarray}
S^\alpha \big(\overrightarrow{\theta^{\alpha}} \big) \equiv -\sum_{G^\alpha\in\mathcal{G}_{N}} P^{\alpha}\big(G^\alpha | \overrightarrow{\theta^{\alpha}}\big) \ln P^{\alpha}\big(G^\alpha| \overrightarrow{\theta^{\alpha}}\big)
\label{entropy_singlelayer}
\end{eqnarray}
is the entropy of the ensemble of monoplex graphs for the individual layer $\alpha$, with $P^{\alpha}\big(G^\alpha | \overrightarrow{\theta^{\alpha}}\big)$ defined as in~\eqref{eq:marginal} and subject to the normalization condition
\begin{equation}
\sum_{G^\alpha\in \mathcal{G}_N}P^{\alpha}\big(G^\alpha|\overrightarrow{\theta^{\alpha}}\big) =1\qquad\forall~\overrightarrow{\theta^{\alpha}}\qquad \alpha=1,M.
\label{eq:normmargin}
\end{equation}\\

At this point, we want to maximize the entropy $\mathcal{S} \big(\overrightarrow{\theta} \big)$, subject to the soft constraint
$\overrightarrow{\mathcal{C}}$, 
to find the functional form of $\mathcal{P}\big(\overrightarrow{G} | \overrightarrow{\theta}\big)$ we are looking for.  Equation~\eqref{entropy} ensures that the maximization of $\mathcal{S} \big(\overrightarrow{\theta} \big)$, subject to~\eqref{eq:constraintmulti}, reduces to the maximization of each single-layer entropy $S^\alpha \big(\overrightarrow{\theta^{\alpha}} \big)$,
subject to 
\begin{equation}
\langle\overrightarrow{C^{\alpha}}\rangle_{\overrightarrow{\theta^{\alpha}}}=\sum_{G^\alpha\in \mathcal{G}_N}P^{\alpha}\big(G^\alpha | \overrightarrow{\theta^{\alpha}}\big)\overrightarrow{C^{\alpha}}(G^\alpha), 
\label{eq:constraintsingle}
\end{equation}
separately.
Therefore the probability $\mathcal{P}\big(\overrightarrow{G} | \overrightarrow{\theta}\big)$ maximizing $\mathcal{S} \big(\overrightarrow{\theta} \big)$ reduces, via~\eqref{prob_multiplex}, to the product of all single-layer probability distributions of the type $P^{\alpha}\big(G^\alpha | \overrightarrow{\theta^{\alpha}}\big)$, each of which should separately maximize the corresponding entropy $S^\alpha \big(\overrightarrow{\theta^{\alpha}} \big)$.\\

The general solution to the problem of maximizing $S^\alpha \big(\overrightarrow{\theta^{\alpha}} \big)$,
subject to~\eqref{eq:constraintsingle}, for single-layer networks is extensively discussed in~\cite{Squartini1} and, in our notation here, leads to the probability distribution
\begin{eqnarray}
P^{\alpha} \big( G^\alpha  | \overrightarrow{\theta^{\alpha}}\big) = \frac{e^{-H^{\alpha}(G^\alpha | \overrightarrow{\theta^{\alpha}}) }}{Z\big(\overrightarrow{\theta^{\alpha}}\big)},
\label{prob}
\end{eqnarray}
where 
\begin{eqnarray}
H^{\alpha} \big(G^\alpha | \overrightarrow{\theta^{\alpha}} \big) = \overrightarrow{\theta^{\alpha}}\cdot \overrightarrow{C^{\alpha}} ( G^\alpha )
\label{hamilt}
\end{eqnarray}
is the \emph{graph Hamiltonian} (the dot indicating a scalar product, i.e. a linear combination of the enforced constraints) and 
\begin{eqnarray}
Z \big(\overrightarrow{\theta^{\alpha}} \big) = \sum_{G^\alpha\in\mathcal{G}_N} e^{- H^{\alpha}(G^\alpha | \overrightarrow{\theta^{\alpha}})}
\label{partit_funct}
\end{eqnarray}
is the \emph{partition function} (representing the normalizing constant for the probability).\\

Equation~\eqref{partit_funct}, and consequently~\eqref{prob}, leads to different explicit functional forms depending on the choice of the constraint(s), i.e. depending on the functional form of $\overrightarrow{C^{\alpha}}(G^\alpha)$. 
In Appendices~\ref{app:DBCM} and~\ref{app:DWCM} we explicitly discuss the cases of the Directed Binary Configuration Model (where the constraints are the in- and out-degrees of all nodes in each layer $\alpha$) and of the Directed Weighted Configuration Model (where the constraints are the in- and out-strenghts of all nodes in each layer $\alpha$), respectively.\\

Once an explicit expression for each $P^{\alpha} \big( G^\alpha | \overrightarrow{\theta^{\alpha}} \big)$ is found, we can use~\eqref{prob_multiplex} to find the final expression for the whole multiplex probability in the null model:
\begin{eqnarray}
\mathcal{P} \big(\overrightarrow{G} |\overrightarrow{\theta}\big) 
= \prod_{\alpha = 1}^{M} \frac{e^{-H^{\alpha}(G^\alpha | \overrightarrow{\theta^{\alpha}}) }}{Z\big(\overrightarrow{\theta^{\alpha}}\big)}
=\frac{e^{-\mathcal{H}(\overrightarrow{G} | \overrightarrow{\theta}) }}{\mathcal{Z}\big(\overrightarrow{\theta}\big)},
\label{prob_final}
\end{eqnarray}
where
\begin{equation}
\mathcal{H}(\overrightarrow{G}|\overrightarrow{\theta})\equiv \sum_{\alpha=1}^M H^{\alpha}(G^\alpha | \overrightarrow{\theta^{\alpha}}) 
\end{equation}
and
\begin{equation}
\mathcal{Z}\big(\overrightarrow{\theta}\big)\equiv\prod_{\alpha=1}^M Z\big(\overrightarrow{\theta^{\alpha}}\big).
\end{equation}
The last three equations rephrase the independence of all layers explicitly. 

\section{Maximum Likelihood Method}
\label{app:maxlike}
The maximization of the entropy is a constrained, \emph{functional} maximization of $\mathcal{S} \big(\overrightarrow{\theta}\big)$ in the space of probability distributions. 
As such, its result is the \emph{functional form} of the maximum-entropy distribution $\mathcal{P} \big(\overrightarrow{G} |\overrightarrow{\theta}\big)$, given by~\eqref{prob_final}, but not its \emph{numerical values}.
In fact, the distribution depends on the whole vector of parameters $\overrightarrow{\theta}$, and any expectation value calculated analytically using the explicit expression of $\mathcal{P} \big(\overrightarrow{G} |\overrightarrow{\theta}\big)$ can only be evaluated numerically after a value of $\overrightarrow{\theta}$ is specified.
This leads to the problem of choosing $\overrightarrow{\theta}$. 
Since we are interested in the case where all layers of the multiplex are independent, choosing a value of $\overrightarrow{\theta}$ reduces to the problem of choosing $\overrightarrow{\theta^{\alpha}}$ separately for each layer.\\

The problem of finding the parameter values of a maximum-entropy model of single-layer networks is solved in the general case in~\cite{Garlaschelli0} using the maximum likelihood principle. 
In our notation here, this solution can be restated as follows.
Let $G^\alpha_*$ denote, among all graphs $G^\alpha\in \mathcal{G}_N$, the particular \emph{empirical} network realized in layer $\alpha$ of the multiplex.
Given $G^\alpha_*$, the log-likelihood function
\begin{eqnarray}
 \mathcal{L}^\alpha\big(\overrightarrow{\theta^{\alpha}} \big) 
\equiv \ln P \big( G_*^{\alpha} | \overrightarrow{\theta^{\alpha}} \big)
\label{likelihood_def}
\end{eqnarray}
represents the log of the probability to generate the empirical graph $G_*^{\alpha} $, given a value of $\overrightarrow{\theta^{\alpha}}$.
The maximum likelihood principle~\cite{Garlaschelli0} states that the optimal choice for $\overrightarrow{\theta^{\alpha}}$ is the one that maximizes the chances to obtain $G_*^{\alpha} $ from the model, i.e. the one that maximizes $\mathcal{L}^\alpha\big(\overrightarrow{\theta^{\alpha}} \big)$.
Let this parameter choice be denoted by $\overrightarrow{\theta_*^{\alpha}}$, where
\begin{equation}
\overrightarrow{\theta_*^{\alpha}}\equiv
\underset{\overrightarrow{\theta^{\alpha}}}{\arg\max} ~\mathcal{L}^\alpha\big(\overrightarrow{\theta^{\alpha}} \big).
\label{maxsam1}
\end{equation}
As a general result~\cite{Garlaschelli0}, the value $\overrightarrow{\theta_*^{\alpha}}$ defined above is such that
\begin{equation}
\langle\overrightarrow{C^{\alpha}}\rangle_{\overrightarrow{\theta_*^{\alpha}}}
=\overrightarrow{C^{\alpha}}(G^\alpha_*),
\label{maxsam2}
\end{equation}
i.e. the expectation value of each constraint coincides with the empirical value measured on the empirical network $G^\alpha_*$.
This is precisely the outcome we desire, given that our ultimate goal is the construction of ensembles of networks with the same numerical value of the constraints as in the real network.\\

From a practical point of view, eqs.~\eqref{maxsam1} and ~\eqref{maxsam2} represent two equivalent ways to determine $\overrightarrow{\theta_*^{\alpha}}$. The former requires the maximization of a scalar function over a $K^\alpha$-dimensional space, while the latter requires the solution of a system of $K^\alpha$ nonlinear coupled equations.
For various choices of the graph ensemble $\mathcal{G}_N$ and of the constraints $\overrightarrow{C^{\alpha}}$ (including those required for our analysis), both approaches are implemented in the MAX$\&$SAM algorithm~\cite{Squartini6}.
More details are given in Appendices \ref{app:DBCM} and \ref{app:DWCM}.
Once the value $\overrightarrow{\theta_*^{\alpha}}$ is found, it is used to find the \emph{numerical value} $ P \big( G^\alpha | \overrightarrow{\theta_*^{\alpha}} \big) $ of the probability of any graph $G^\alpha\in\mathcal{G}_N$.
So, while the maximization of the entropy generates the functional form of the graph probability, the maximization of the likelihood fixes its numerical values. 
If $X^\alpha$ denotes any single-layer structural property $X$ of interest, the above procedure allows us to evaluate the expected value
\begin{equation}
\langle X^\alpha\rangle \equiv
\langle X^\alpha\rangle_{\overrightarrow{\theta_*^{\alpha}}}
=\sum_{G^\alpha\in\mathcal{G}_N}P \big( G^\alpha | \overrightarrow{\theta_*^{\alpha}} \big)X^\alpha(G^\alpha)
\label{eq:X}
\end{equation}
(and similarly the standard deviation) of $X^\alpha$ explicitly over the desired ensemble~\cite{Garlaschelli0, Squartini1, Squartini6}.
For many properties of interest, the expected value~\eqref{eq:X} can be calculated analytically given the explicit expression of $P \big( G^\alpha | \overrightarrow{\theta_*^{\alpha}}\big)$, without the need to sample the graph ensemble explicitly~\cite{Squartini1}. 
For more complicated properties, one can instead use the knowledge of $P \big( G^\alpha | \overrightarrow{\theta_*^{\alpha}}\big)$ to sample graphs from the ensemble in an unbiased way and then calculate expectations as sample averages~\cite{Squartini6}.\\

The multiplexity and multireciprocity metrics introduced in sec.\ref{sec:defs} are not single-layer properties like $X^\alpha$, as they require measurements on multiple layers simultaneously. 
Using eq.~\eqref{prob_multiplex}, we therefore need to generalize eq.~\eqref{eq:X} to the case of an arbitrary multiplex quantity $\mathcal{X}$, evaluating to $\mathcal{X}(\overrightarrow{G})$ on a specific multiplex $\overrightarrow{G}\in\mathcal{G}_N^M$, as follows:
\begin{equation}
\langle \mathcal{X}\rangle \equiv
\langle \mathcal{X}\rangle_{\overrightarrow{\theta_*}}
=\sum_{\overrightarrow{G}\in\mathcal{G}_N^M}\mathcal{P} \big( \overrightarrow{G} | \overrightarrow{\theta_*} \big)\mathcal{X}(\overrightarrow{G})
\label{eq:XX}
\end{equation}
where $\overrightarrow{\theta_*}=(\overrightarrow{\theta^\alpha_*})_{\alpha=1}^M$ 
contains the Langrange multipliers~\eqref{maxsam1} for all layers and  $\overrightarrow{G_*}=(G^\alpha_*)_{\alpha=1}^M\in\mathcal{G}_N^M$ denotes the whole empirical multiplex.
Both the expected values and the standard deviations of multiplexity and multireciprocity can be calculated explicitly, and we will therefore follow the analytical approach, which is exact and faster than the sampling approach  (see Appendices \ref{app:DBCM} and \ref{app:DWCM}).\\

From a computational point of view, the above canonical approach based on soft constraints has many benefits with respect to the microcanonical approach with hard constraints~\cite{Squartini1,Squartini6}. 
Indeed, the microcanonical approach cannot be controlled analytically, and necessarily requires sampling many randomized multiplexes explicitly from the ensemble.
Generating even only a single randomized multiplex requires the iteration of many random constraint-preserving `rewiring moves', which is computationally costly.
Such procedure must be repeated several times, to produce a large sample of $ R $ randomized multiplexes, on each of which any topological property $X$ of interest has to be calculated. Finally, a sample average should be performed to obtain an estimate of $\langle X\rangle$.\\

For instance, on single-layer networks with constrained degree sequence one should iterate the so-called `local rewiring algorithm'~\cite{Maslov} that preserves the degrees while randomizing the network. On a monoplex network with $L$ links, the above approach would require a computational time of order $ O(L)$, only to generate a single realization of the randomized network.
On such realization, one would then need to measure $X$ (for instance the monoplex reciprocity), which would require a certain time $T_X$. 
The total time needed for a single realization would therefore be $T_X+O(L)$, and for all realizations $R\cdot T_X+O(R\cdot L)$.\\

In a multiplex network with $M$ layers, the corresponding time required to generate a single randomized multiplex would in principle be of order $ O(\sum_{\alpha=1}^M L^\alpha) $, where $ L^{\alpha} $ is the number of links in the $ \alpha$-th layer. 
However, if layers are independent in the null model, the randomization could (if computational resource allows) be run in parallel on the different layers, thus reducing the above time to $O(\bar{L})$ where $\bar{L}$ is the average number of links per layer, which does not scale with $M$. 
However, the calculation of multiplex quantities $X$ (e.g. the multireciprocity) which would require a time $T_X$ for a single layer (e.g. the monoplex reciprocity) would now need to be iterated for each pair of layers, thus requiring a time $O(T_X\cdot M^2)$.
In total, this means that the total microcanonical computational time for a multiplex is $T_\textrm{mic}=O(R\cdot T_X\cdot M^2)+O(R\cdot L)$, before carrying out the final sample averages.\\

By contrast, our canonical approach does not require the sampling of any multiplex. For individual layers, the calculation of the expected value of most properties of interest basically requires replacing the adjacency matrix of the network with the corresponding expected matrix (or more complicated replacements that in any case require a comparable calculation time).
Therefore calculating the expected value $\langle X\rangle$ takes the same time $T_X$ that it would take for the empirical property $X$ to be calculated on the real system. The same holds true for the entire multiplex. 
Therefore the total canonical time needed is
$T_\textrm{can}=O( T_X\cdot M^2)+T_{\mathcal{L}}$, where $ T_{\mathcal{L}}$ is the one-off time required to preliminary maximize the likelihood (possibly of each layer in parallel) defined in~(\ref{likelihood_def}). 

As already mentioned above, the time $ T_{\mathcal{L}}$ required to maximize the likelihood function can be 
proxied by the time required to solve a system of coupled, non-linear 
equations ( $2N$ equations in the case of directed networks, as shown below). However, since such systems can 
be further simplified by rewriting them only in terms of the sequences of distinct 
directed degrees/strengths (which are always less than $2N$), the 
computational time drops to the order of seconds or minutes (depending on the chosen constraints) for each layer. 
Moreover, further analyses on synthetic networks have shown that this time scales roughly quadratically with the number of nodes; 
it is anyway considerably shorter than the corresponding total microcanonical time $T_\textrm{mic}$ estimated above.\\

Besides the computational advantages described above, the canonical approach has the statistical advantage of being a truly unbiased method~\cite{Squartini6}, in the sense that its maximum-entropy nature implies that no preference is given to specific graph configurations, other than on the basis of the enforced constraints.
So unbiasedness is ensured by the maximum degree of randomness encoded in the graph probability, given the constraints.
By constrast, microcanonical approaches are not guaranteed to ensure the same property.
In the microcanonical case, unbiasedness means that the realizations of the network should be sampled uniformly (i.e. with exactly the same probability) from the whole set of configurations compatible with the constraints. 
Ensuring uniform sampling is highly nontrivial and often impossible.
For instance, in the case of graphs with fixed degree sequence, it can be proved that the local rewiring algorithm is biased, as it preferentially samples configurations that are `close' to the empirical one~\cite{Roberts}.
The authors showed that it is in principle possible to remove this bias, by calculating the so-called `mobility' function (which is a quantity that depends on the current configuration being randomized) and accepting the `next' randomized configurations with a probability that depends on the mobility itself. 
This requirement further increases, and by a large extent, the already heavy computational requirements of the microcanonical approach, because the mobility should be continuously recalculated during the randomization process.

\section{Directed binary configuration model}
\label{app:DBCM}
In this Appendix we explicitly discuss the DBCM model~\cite{Squartini1,Squartini6}, obtained through the maximum entropy and maximum likelihood methods in the specific case where $\mathcal{G}_N$ contains all binary directed graphs with $N$ nodes and $\overrightarrow{C^{\alpha}}$ is a vector of dimension $K^\alpha=2N$ containing the out-degree $k_{i}^{out}$ and the in-degree $k_{i}^{in}$ of all nodes ($i=1,N$). 
Correspondingly, the $2N$-dimensional vector $\overrightarrow{\theta^{\alpha}}$ contains the associated Lagrange multipliers $\phi^{\alpha}_i$ and $\chi^{\alpha}_i$ for all nodes.
Note that we enforce the in- and out-degree sequences on all layers, which means that, as a function, $\overrightarrow{C^{\alpha}}=
(\overrightarrow{k^{out}},\overrightarrow{k^{in}})$ is the same for all $\alpha$. 
However, the numerical values of the degrees in different layers will in general be different, i.e. $\overrightarrow{C}(G^\alpha)\ne \overrightarrow{C}(G^\beta)$ for $\alpha\ne\beta$, thus $\overrightarrow{\theta^{\alpha}}=(\overrightarrow{\phi^{\alpha}}, \overrightarrow{\chi^{\alpha}})$ must still depend on $\alpha$ explictly.\\

For single-layer networks, this model is discussed e.g. in~\cite{Squartini1,Squartini6}.
Here we simply summarize the main steps leading to the final expressions for the expected binary multiplexity and binary reciprocity.
Using the notation introduced in Sec.~\ref{sec:defs} and in Appendix~\ref{app:maxent}, the single-layer Hamiltonian~\eqref{hamilt} reads
\begin{eqnarray}
H \big(G^{\alpha} | \overrightarrow{\phi^{\alpha}}, \overrightarrow{\chi^{\alpha}} \big) & = & \overrightarrow{\phi^{\alpha}}\cdot\overrightarrow{ k^{out}}(G^\alpha) + \overrightarrow{\chi^{\alpha}}\cdot\overrightarrow{ k^{in}}(G^\alpha)  = \nonumber \\
& = & \sum_{i=1}^{N} \left[ \phi_i^{\alpha} k_{i}^{out}(G^\alpha) + \chi_i^{\alpha} k_{i}^{in}(G^\alpha) \right] = \nonumber \\ & = & \sum_{i=1}^N\sum_{j\ne i}\big( \phi_i^{\alpha} + \chi_j^{\alpha} \big) a_{ij}^{\alpha}
\label{hamilt_bin}
\end{eqnarray}
and the partition function~\eqref{partit_funct} can be calculated as:
\begin{eqnarray}
Z \big(\overrightarrow{\phi^{\alpha}}, \overrightarrow{\chi^{\alpha}} \big) & = & \prod_{i=1}^N\prod_{j \neq i} \big( 1 + e^{-\phi_i^{\alpha} - \chi_j^{\alpha}} \big)  \nonumber \\
& = & \prod_{i=1}^N\prod_{j\ne i} \big( 1 + x_i^{\alpha} y_j^{\alpha} \big),
\label{partition_bin}
\end{eqnarray}
where we have set $x_i^{\alpha}\equiv e^{-\phi_i^{\alpha}}$ and $y_i^{\alpha}\equiv e^{-\chi_i^{\alpha}}$. 
This implies that the probability~\eqref{prob} can be written explicitly as
\begin{eqnarray}
P^\alpha \big(G^\alpha|\overrightarrow{\phi^{\alpha}}, \overrightarrow{\chi^{\alpha}} \big) & = & \prod_{i=1}^N\prod_{j \neq i} \frac{ (x_i^{\alpha} y_j^{\alpha})^{a_{ij}^\alpha}}{ 1 + x_i^{\alpha} y_j^{\alpha}}\nonumber\\
& = & \prod_{i=1}^N\prod_{j\ne i} (p_{ij}^{\alpha} )^{a_{ij}^\alpha}(1-p_{ij}^{\alpha} )^{1-a_{ij}^\alpha},
\label{ppp}
\end{eqnarray}
where
\begin{eqnarray}
p_{ij}^{\alpha} = \frac{x_i^{\alpha} y_j^{\alpha}}{1 + x_i^{\alpha} y_j^{\alpha}}
\label{app_def_prob}
\end{eqnarray}
is the probability of a directed link from $i$ to $j$ in layer $\alpha$.
Equation~\eqref{ppp} shows that the random variable $a_{ij}^\alpha$ is drawn, for all $i\ne j$, from a Bernoulli distribution with success probability $p_{ij}^\alpha$, thus leading to eq.~\eqref{eq:Bernoulli}.\\

Given the real-world multiplex $\overrightarrow{G}_*=(G^\alpha_*)_{\alpha=1}^M$, the single-layer log-likelihood function~\eqref{likelihood_def} to be maximized is then given by
\begin{eqnarray}
 \mathcal{L} \big(\overrightarrow{x^{\alpha}}, \overrightarrow{y^{\alpha}} \big) & = & \sum_{i=1}^{N} \left[ k_{i}^{out}(G^\alpha_*)  \ln x_i^{\alpha} + k_{i}^{in}(G_*^\alpha)  \ln y_i^{\alpha} \right] + \nonumber \\
  &  & -\sum_{i=1}^N\sum_{j\ne i}\ln \big( 1 + x_i^{\alpha} y_j^{\alpha} \big),
\label{likelihood_bin}
\end{eqnarray}
and the equivalent set of $ 2N $ coupled nonlinear equations~\eqref{maxsam2} to be solved is
\begin{eqnarray}
\sum_{j \neq i} \frac{x_i^{\alpha} y_j^{\alpha}}{1 + x_i^{\alpha} y_j^{\alpha}} &=& k_{i}^{out}(G_*^\alpha) \qquad\forall i = 1, N\\
\sum_{j \neq i} \frac{x_j^{\alpha} y_i^{\alpha}}{1 + x_j^{\alpha} y_i^{\alpha}} &=& k_{i}^{in}(G_*^\alpha)\qquad \forall i = 1,N.
\end{eqnarray}
Once found, the values of $ \{ x_{i}^{\alpha} \} $ and $ \{ y_{i}^{\alpha} \} $ providing the unique solution to the above problem can be put back in eqs.~\eqref{ppp} and~\eqref{app_def_prob}, allowing us to analytically calculate the expected values $\langle\cdot\rangle_\textrm{DBCM}$ of the quantities of interest via the corresponding probabilities $p_{ij}^\alpha$ (where for simplicity we drop the asterisk indicating that $p_{ij}^\alpha$ is evaluated at the specific values that maximize the likelihood).\\

In particular, we can calculate the rescaled metrics defined in eqs.~\eqref{def_mu_bin} as follows.
First of all, since in the DBCM the in- and out-degrees of all nodes in all layers are equal to their expected values, we necessarily have $\langle L^\alpha\rangle_\textrm{DBCM}=L^\alpha_*$ for all $\alpha$, where $L^\alpha_*\equiv L^\alpha(G^\alpha_*)$ is the number of links of the observed, layer-specific graph $G^\alpha_*$.
This means that $L^\alpha$ is a constrained quantity, and we therefore expect the denominators of eqs.~\eqref{m_bin} to fluctuate around their expected values $L^\alpha_*+L^\beta_*$ much less than how the  numerators fluctuate around the corresponding expected values.
We therefore approximate the expected values of eqs.~\eqref{m_bin} as follows:
\begin{subequations}
\begin{align}
\langle m^{\alpha, \beta}_b\rangle_\textrm{DBCM}= \frac{2 \langle L^{\alpha\rightrightarrows\beta}\rangle_\textrm{DBCM}}{L_*^{\alpha} + L_*^{\beta}}&\qquad(\alpha\ne\beta),\\ 
\langle r^{\alpha, \beta}_b\rangle_\textrm{DBCM} = \frac{2 \langle L^{\alpha\rightleftarrows\beta}\rangle_\textrm{DBCM}}{L_*^{\alpha} + L_*^{\beta}}.&
\end{align}
\label{expm_bin}
\end{subequations}
Consequently,
\begin{eqnarray}
\mu^{\alpha, \beta}_b &=& 
\frac{2L_*^{\alpha\rightrightarrows\beta}-2\langle L^{\alpha\rightrightarrows\beta}\rangle_{\rm DBCM}}
{L^\alpha_*+L^\beta_* - 2 \langle L^{\alpha\rightrightarrows\beta} \rangle_\textrm{DBCM}}\qquad(\alpha\ne\beta),\nonumber\\
\rho^{\alpha, \beta}_b &=& 
\frac{2L_*^{\alpha\rightleftarrows\beta}-2\langle L^{\alpha\rightleftarrows\beta}\rangle_{\rm DBCM}}
{L^\alpha_*+L^\beta_* - 2 \langle L^{\alpha\rightleftarrows\beta} \rangle_\textrm{DBCM}}\nonumber.
\end{eqnarray}
Since $a_{ij}^\alpha$ and $a_{ij}^\beta$ (for $\beta\ne \alpha$), and similarly $a_{ij}^\alpha$ and $a_{ji}^\beta$ (for any $\beta$), are independently drawn from two Bernoulli distributions, the expected values of $\min \{ a_{ij}^{\alpha}, a_{ij}^{\beta}\}$ (with $\beta\ne\alpha$) and $\min \{ a_{ij}^{\alpha}, a_{ji}^{\beta}\}$ are easily calculated as
\begin{subequations}
\begin{align}
\langle \min \{ a_{ij}^{\alpha}, a_{ij}^{\beta} \} \rangle_{\rm DBCM} =p_{ij}^{\alpha}p_{ij}^{\beta}&\qquad (\alpha\ne\beta), \\ 
\langle \min \{ a_{ij}^{\alpha}, a_{ji}^{\beta} \} \rangle_{\rm DBCM} = p_{ij}^{\alpha} p_{ji}^{\beta}. &
\end{align}
\label{app_min_bin}
\end{subequations}
Therefore the final expressions for the transformed multiplexity and multireciprocity are:
\begin{subequations}
\begin{align}
\mu^{\alpha, \beta}_b = \frac{2L_*^{\alpha\rightrightarrows\beta}-2 \sum_i\sum_{j\ne i} p_{ij}^{\alpha} p_{ij}^{\beta}}{L_*^{\alpha}+L_*^{\beta}-2\sum_i\sum_{j\ne i}  p_{ij}^{\alpha} p_{ij}^{\beta}}&\quad (\alpha\ne\beta)\\
\rho^{\alpha, \beta}_b = \frac{2L_*^{\alpha\rightleftarrows\beta}-2 \sum_i\sum_{j\ne i} p_{ij}^{\alpha} p_{ji}^{\beta}}{L_*^{\alpha}+L_*^{\beta}-2\sum_i\sum_{j\ne i}  p_{ij}^{\alpha} p_{ji}^{\beta}},
\end{align}
\label{app_rho_bin_spec}
\end{subequations}
where the probabilities are defined according to Eq.~(\ref{app_def_prob}).\\

Similarly, we need to calculate the $z$-scores defined in eqs.~\eqref{zScores}.
To do this, we need to calculate the standard deviations of $m_b^{\alpha,\beta}$ and $r_b^{\alpha,\beta}$ at the denominator of the $z$-scores.
Neglecting again the fluctuations of the constrained quantities $L^\alpha$ and $L^\beta$ around their average values (with respect to the fluctuations of the unconstrained quantities), and since all pairs of nodes are independent, we calculate the variances of $m_b^{\alpha,\beta}$ and $r_b^{\alpha,\beta}$ in a way similar to what we did for the expressions in eq.~\eqref{expm_bin}:
\begin{subequations}
\begin{align}
\textrm{Var}[ m^{\alpha, \beta}_b]= \frac{4  \sum_i\sum_{j \neq i} \textrm{Var}[ \min \{ a_{ij}^{\alpha}, a_{ij}^{\beta}\}]}{(L_*^{\alpha} + L_*^{\beta})^2}&\qquad(\alpha\ne\beta),\nonumber\\ 
\textrm{Var}[ r^{\alpha, \beta}_b]= \frac{4  \sum_i\sum_{j \neq i} \textrm{Var}[ \min \{ a_{ij}^{\alpha}, a_{ji}^{\beta}\}]}{(L_*^{\alpha} + L_*^{\beta})^2}&.\nonumber
\end{align}
\label{expm_bin}
\end{subequations}
Now we note that the minimum of two $0/1$ quantities is also a $0/1$ quantity.
This implies that the square of the minimum is equal to the minimum itself, and that the expected square of the minimum is equal to the expected value of the minimum. In formulas:
\begin{eqnarray}
\langle {\min}^2 \{ a_{ij}^{\alpha}, a_{ij}^{\beta} \} \rangle_{\rm DBCM}&=& p_{ij}^{\alpha} p_{ij}^{\beta}\qquad(\alpha\ne\beta),\\
\langle {\min}^2 \{ a_{ij}^{\alpha}, a_{ji}^{\beta} \} \rangle_{\rm DBCM}&=& p_{ij}^{\alpha} p_{ji}^{\beta}.
\label{app_min_quadro}
\end{eqnarray}
It then follows that the variance of the minimum is 
\begin{eqnarray}
\textrm{Var} \big[\min \{ a_{ij}^{\alpha}, a_{ij}^{\beta}\} \big] &=&p_{ij}^{\alpha} p_{ij}^{\beta}(1-p_{ij}^{\alpha} p_{ij}^{\beta})\qquad(\alpha\ne\beta),\nonumber\\
\textrm{Var} \big[\min \{ a_{ij}^{\alpha}, a_{ji}^{\beta}\} \big] &=&p_{ij}^{\alpha} p_{ji}^{\beta}(1-p_{ij}^{\alpha} p_{ji}^{\beta}).\nonumber
\end{eqnarray}
Putting these expressions into those for $\textrm{Var}[ m^{\alpha, \beta}_b]$ and $\textrm{Var}[ r^{\alpha, \beta}_b]$, and taking the square root to obtain the standard deviations, we finally arrive at the explicit calculation of the $z$-scores:
\begin{eqnarray}
z \big( m^{\alpha, \beta}_b \big) &=& \frac{L_*^{\alpha\rightrightarrows\beta}-\sum_i\sum_{j\ne i} p_{ij}^{\alpha}p_{ij}^{\beta}}{\sqrt{\sum_i\sum_{j\ne i} p_{ij}^{\alpha}p_{ij}^{\beta}(1 - p_{ij}^{\alpha}p_{ij}^{\beta})}}\quad(\alpha\ne\beta)\nonumber\\
z \big( r^{\alpha, \beta}_b \big) &=& \frac{L_*^{\alpha\rightleftarrows\beta}-\sum_i\sum_{j\ne i}  p_{ij}^{\alpha}p_{ji}^{\beta}}{\sqrt{\sum_i\sum_{j\ne i} p_{ij}^{\alpha}p_{ji}^{\beta}(1 - p_{ij}^{\alpha}p_{ji}^{\beta})}}\nonumber
\end{eqnarray}
From a direct comparison between the above equations and Eqs.~(\ref{app_rho_bin_spec}), we immediately observe the sign concordance reported in the main text.

\section{Directed weighted configuration model}
\label{app:DWCM}
Here we consider the DWCM model~\cite{Squartini1,Squartini6}, obtained when $\mathcal{G}_N$ contains all weighted directed graphs (with non-negative integer edge weights) with $N$ nodes and $\overrightarrow{C^{\alpha}}$ is a vector of dimension $K^\alpha=2N$ containing the out-strength $s_{i}^{out}$ and the in-strength $s_{i}^{in}$ of all nodes ($i=1,N$). 
The $2N$-dimensional vector $\overrightarrow{\theta^{\alpha}}$ contains the associated Lagrange multipliers $\phi^{\alpha}_i$ and $\chi^{\alpha}_i$.
As for the DBCM, $\overrightarrow{C^{\alpha}}=
(\overrightarrow{s^{out}},\overrightarrow{s^{in}})$ is the same function for all $\alpha$. 
However, the numerical values  $\overrightarrow{\theta^{\alpha}}=(\overrightarrow{\phi^{\alpha}}, \overrightarrow{\chi^{\alpha}})$ still depend on $\alpha$.\\

For single-layer networks (see e.g. discussion in~\cite{Squartini1,Squartini6}), the Hamiltonian~\eqref{hamilt} reads
\begin{eqnarray}
H \big(G^{\alpha} | \overrightarrow{\phi^{\alpha}}, \overrightarrow{\chi^{\alpha}} \big) & = & \overrightarrow{\phi^{\alpha}}\cdot\overrightarrow{ s^{out}}(G^\alpha) + \overrightarrow{\chi^{\alpha}}\cdot\overrightarrow{ s^{in}}(G^\alpha)  = \nonumber \\
& = & \sum_{i=1}^{N} \left[ \phi_i^{\alpha} s_{i}^{out}(G^\alpha) + \chi_i^{\alpha} s_{i}^{in}(G^\alpha) \right] = \nonumber \\ & = & \sum_{i=1}^N\sum_{j\ne i}\big( \phi_i^{\alpha} + \chi_j^{\alpha} \big) w_{ij}^{\alpha}
\label{hamilt_wei}
\end{eqnarray}
and the partition function~\eqref{partit_funct} can be calculated as:
\begin{eqnarray}
Z \big(\overrightarrow{\phi^{\alpha}}, \overrightarrow{\chi^{\alpha}} \big) & = & \prod_{i=1}^N\prod_{j \neq i} \big( 1 - e^{-\phi_i^{\alpha} - \chi_j^{\alpha}} \big)^{-1}  \nonumber \\
& = & \prod_{i=1}^N\prod_{j\ne i} \big( 1 - x_i^{\alpha} y_j^{\alpha} \big)^{-1},
\label{partition_wei}
\end{eqnarray}
where we have set $x_i^{\alpha}\equiv e^{-\phi_i^{\alpha}}$ and $y_i^{\alpha}\equiv e^{-\chi_i^{\alpha}}$. 
This implies that the probability~\eqref{prob} can be written as
\begin{eqnarray}
P^\alpha \big(G^\alpha|\overrightarrow{\phi^{\alpha}}, \overrightarrow{\chi^{\alpha}} \big) & = & \prod_{i=1}^N\prod_{j \neq i}  (x_i^{\alpha} y_j^{\alpha})^{w_{ij}^\alpha}( 1- x_i^{\alpha} y_j^{\alpha})\nonumber\\
& = & \prod_{i=1}^N\prod_{j\ne i} (p_{ij}^{\alpha} )^{w_{ij}^\alpha}(1-p_{ij}^{\alpha} ),
\label{ppp2}
\end{eqnarray}
where
\begin{eqnarray}
p_{ij}^{\alpha} = {x_i^{\alpha} y_j^{\alpha}}
\label{app_def_prob2}
\end{eqnarray}
denotes again the probability that a directed link (of any positive weight) from node $i$ to node $j$ is realized in layer $\alpha$.
Equation~\eqref{ppp2} gives the interpretation of $w_{ij}^\alpha$ as a geometrically distributed variable, constructed as the iteration of many random events, each defined as incrementing $w_{ij}^\alpha$ by one, starting from $w_{ij}^\alpha=0$.
In this interpretation, $p_{ij}^{\alpha}$ is the elementary probability of a `success' event, and the probability that $w_{ij}^\alpha=w$ coincides with the probability $ (p_{ij}^{\alpha} )^{w}(1-p_{ij}^{\alpha} )$ of having $w$ consecutive successes followed by one failure.
This leads to the geometric distribution in eq.~\eqref{eq:Geometric}.\\

The single-layer log-likelihood function~\eqref{likelihood_def} to be maximized is now given by
\begin{eqnarray}
 \mathcal{L} \big(\overrightarrow{x^{\alpha}}, \overrightarrow{y^{\alpha}} \big) & = & \sum_{i=1}^{N} \left[ s_{i}^{out}(G^\alpha_*)  \ln x_i^{\alpha} + s_{i}^{in}(G_*^\alpha)  \ln y_i^{\alpha} \right] + \nonumber \\
  &  & +\sum_{i=1}^N\sum_{j\ne i}\ln \big( 1 - x_i^{\alpha} y_j^{\alpha} \big),
\label{likelihood_wei}
\end{eqnarray}
and the corresponding equations~\eqref{maxsam2} are
\begin{eqnarray}
\sum_{j \neq i} \frac{x_i^{\alpha} y_j^{\alpha}}{1 - x_i^{\alpha} y_j^{\alpha}} &=& s_{i}^{out}(G_*^\alpha) \qquad\forall i = 1, N\\
\sum_{j \neq i} \frac{x_j^{\alpha} y_i^{\alpha}}{1 - x_j^{\alpha} y_i^{\alpha}} &=& s_{i}^{in}(G_*^\alpha)\qquad \forall i = 1,N.
\end{eqnarray}
The expected values $\langle\cdot\rangle_\textrm{DWCM}$ of the relevant quantities can be found through eqs.~\eqref{ppp2} and~\eqref{app_def_prob2}, evaluated at the values of $ \{ x_{i}^{\alpha} \} $ and $ \{ y_{i}^{\alpha} \} $ that solve the above problem (again, in what follows we drop the asterisk indicating that $p_{ij}^\alpha$ is evaluated at the specific values that maximize the likelihood).\\

We start with the calculation of the expected values of the multiplexity and multireciprocity metrics defined in eq.~\eqref{m_wei}.
In analogy with what we did for the DBCM, we expect the (constrained) denominators of eqs.~\eqref{m_wei} to fluctuate much less than the (unconstrained) numerators and we therefore replace the denominators with their expected values $W^\alpha_*+W^\beta_*$.
We therefore write
\begin{subequations}
\begin{align}
\langle m^{\alpha, \beta}_w\rangle_\textrm{DWCM}= \frac{2   \langle W^{\alpha\rightrightarrows\beta}\rangle_\textrm{DWCM}}{W_*^{\alpha} + W_*^{\beta}},&\\ 
\langle r^{\alpha, \beta}_w\rangle_\textrm{DWCM} = \frac{2 \langle W^{\alpha\rightleftarrows\beta}\rangle_\textrm{DWCM}}{W_*^{\alpha} + W_*^{\beta}} &
\end{align}
\label{expm_wei}
\end{subequations}
and
\begin{eqnarray}
\mu^{\alpha, \beta}_w &=& 
\frac{2W_*^{\alpha\rightrightarrows\beta}-2\langle W^{\alpha\rightrightarrows\beta}\rangle_{\rm DWCM}}
{W^\alpha_*+W^\beta_* - 2 \langle W^{\alpha\rightrightarrows\beta} \rangle_\textrm{DWCM}}\qquad(\alpha\ne\beta),\nonumber\\
\rho^{\alpha, \beta}_w &=& 
\frac{2W_*^{\alpha\rightleftarrows\beta}-2\langle W^{\alpha\rightleftarrows\beta}\rangle_{\rm DWCM}}
{W^\alpha_*+W^\beta_* - 2 \langle W^{\alpha\rightleftarrows\beta} \rangle_\textrm{DWCM}}\nonumber.
\end{eqnarray}
Since $w_{ij}^\alpha$ and $w_{ij}^\beta$ (for $\beta\ne \alpha$), and similarly $w_{ij}^\alpha$ and $w_{ji}^\beta$ (for any $\beta$), are independenlty drawn from two geometric distributions, the expected values of $\min \{ w_{ij}^{\alpha}, w_{ij}^{\beta}\}$ (with $\beta\ne\alpha$) and $\min \{ w_{ij}^{\alpha}, w_{ji}^{\beta}\}$ are easily calculated as
\begin{subequations}
\begin{align}
\langle \min \{ a_{ij}^{\alpha}, a_{ij}^{\beta} \} \rangle_{\rm DBCM} =\frac{p_{ij}^{\alpha} p_{ij}^{\beta}}{1 - p_{ij}^{\alpha} p_{ij}^{\beta}}&\qquad (\alpha\ne\beta), \\ 
\langle \min \{ a_{ij}^{\alpha}, a_{ji}^{\beta} \} \rangle_{\rm DBCM} =\frac{p_{ij}^{\alpha} p_{ji}^{\beta}}{1 - p_{ij}^{\alpha} p_{ji}^{\beta}}. &
\end{align}
\label{app_min_wei}
\end{subequations}
Therefore the transformed multiplexity and multireciprocity read
\begin{subequations}
\begin{align}
\mu^{\alpha, \beta}_w = \frac{2W_*^{\alpha\rightrightarrows\beta}-2 \sum_i\sum_{j\ne i} \frac{p_{ij}^{\alpha} p_{ij}^{\beta}}{1 - p_{ij}^{\alpha}p_{ij}^{\beta}}}{W_*^{\alpha}+W_*^{\beta}-2\sum_i\sum_{j\ne i}  \frac{p_{ij}^{\alpha} p_{ij}^{\beta}}{1 - p_{ij}^{\alpha} p_{ij}^{\beta}}}&\quad (\alpha\ne\beta)\\
\rho^{\alpha, \beta}_w = \frac{2W_*^{\alpha\rightleftarrows\beta}-2 \sum_i\sum_{j\ne i} \frac{p_{ij}^{\alpha} p_{ji}^{\beta}}{1 - p_{ij}^{\alpha} p_{ji}^{\beta}}}{W_*^{\alpha}+W_*^{\beta}-2\sum_i\sum_{j\ne i}  \frac{p_{ij}^{\alpha} p_{ji}^{\beta}}{1 - p_{ij}^{\alpha} p_{ji}^{\beta}}},
\end{align}
\label{app_rho_wei_spec}
\end{subequations}
where the probabilities are defined in Eq.~(\ref{app_def_prob2}).\\

We then calculate the $z$-scores.
Following an argument similar to the binary case, we write
\begin{subequations}
\begin{align}
\textrm{Var}[ m^{\alpha, \beta}_w]= \frac{4  \sum_i\sum_{j \neq i} \textrm{Var}[ \min \{ w_{ij}^{\alpha}, w_{ij}^{\beta}\}]}{(W_*^{\alpha} + W_*^{\beta})^2}&\qquad(\alpha\ne\beta),\nonumber\\ 
\textrm{Var}[ r^{\alpha, \beta}_w]= \frac{4  \sum_i\sum_{j \neq i} \textrm{Var}[ \min \{ w_{ij}^{\alpha}, w_{ji}^{\beta}\}]}{(W_*^{\alpha} + W_*^{\beta})^2}&.\nonumber
\end{align}
\label{expm_wei}
\end{subequations}
After calculating the variance of the minimum of two geometrically distributed random variables, we get
 \begin{eqnarray}
\textrm{Var} \big[\min \{ w_{ij}^{\alpha}, w_{ij}^{\beta}\} \big] &=&\frac{p_{ij}^{\alpha} p_{ij}^{\beta} }{\big(1 - p_{ij}^{\alpha} p_{ij}^{\beta} \big)^2}\qquad(\alpha\ne\beta),\nonumber\\
\textrm{Var} \big[\min \{ a_{ij}^{\alpha}, a_{ji}^{\beta}\} \big] &=&\frac{p_{ij}^{\alpha} p_{ji}^{\beta} }{\big(1 - p_{ij}^{\alpha} p_{ji}^{\beta} \big)^2}.\nonumber
\end{eqnarray}
Combining all the relevant expressions together, we get for the $z$-scores:
\begin{eqnarray}
z \big( m^{\alpha, \beta}_w \big) &=& \frac{W_*^{\alpha\rightrightarrows\beta}-\sum_i\sum_{j\ne i} \frac{p_{ij}^{\alpha} p_{ij}^{\beta} }{1 - p_{ij}^{\alpha} p_{ij}^{\beta} }  }{\sqrt{\sum_i\sum_{j\ne i} \frac{p_{ij}^{\alpha} p_{ij}^{\beta} }{\big(1 - p_{ij}^{\alpha} p_{ij}^{\beta} \big)^2}}}\quad(\alpha\ne\beta)\nonumber\\
z \big( r^{\alpha, \beta}_w \big) &=& \frac{W_*^{\alpha\rightleftarrows\beta}-\sum_i\sum_{j\ne i} \frac{p_{ij}^{\alpha} p_{ji}^{\beta} }{1 - p_{ij}^{\alpha} p_{ji}^{\beta} }  }{\sqrt{\sum_i\sum_{j\ne i} \frac{p_{ij}^{\alpha} p_{ji}^{\beta} }{\big(1 - p_{ij}^{\alpha} p_{ji}^{\beta} \big)^2}}}\nonumber
\end{eqnarray}
Comparing with Eqs.~(\ref{app_rho_wei_spec}), we confirm the concordance of the sings.



\bibliography{Gemmetto_et_al_pre_references_august2016.bib}

\end{document}